\documentclass[twocolumn]{aastex62}
\usepackage[T1]{fontenc}
\usepackage[utf8]{inputenc}
\usepackage[english]{babel}
\usepackage{amssymb}
\usepackage{amsmath}
\usepackage{textgreek}

\newcommand\tsup[1]{\textsuperscript{#1}}

\newcommand\supn[1]{\tsup{\ensuremath{#1}}}

\makeatletter
\newcommand\DeclareUnit[2]{%
    \@namedef{#1}{\@ifnextchar[{\csname @with@#1\endcsname}{\csname @without@#1\endcsname}}%
    \@namedef{@with@#1}[##1]{\text{#2\supn{##1}}}%
    \@namedef{@without@#1}{\text{#2}}%
}%
\makeatother

\DeclareUnit{km}{\text{km}}
\DeclareUnit{m}{\text{m}}
\DeclareUnit{cm}{\text{cm}}
\DeclareUnit{mm}{\text{mm}}
\DeclareUnit{um}{\text{\textmu m}}
\DeclareUnit{s}{\text{s}}
\DeclareUnit{hr}{\text{hr}}
\DeclareUnit{kg}{\text{kg}}
\DeclareUnit{g}{\text{g}}
\DeclareUnit{pc}{\text{pc}}
\DeclareUnit{au}{\text{AU}}
\DeclareUnit{Pa}{\text{Pa}}
\DeclareUnit{K}{\text{K}}
\DeclareUnit{erg}{\text{erg}}

\newcommand\Gpc{\mathrm{Gpc}}
\newcommand\Mpc{\mathrm{Mpc}}
\newcommand\yr{\mathrm{yr}}

\begin{document}
\title{Please repeat: Strong lensing of gravitational waves as a probe of compact binary and galaxy populations}

\author{Fei Xu}
\correspondingauthor{Fei Xu}
\email{feixu@uchicago.edu}
\affiliation{Department of Astronomy and Astrophysics, University of Chicago, 5640 South Ellis Ave., Chicago, IL 60637}

\author{Jose Mar\'ia Ezquiaga}
\altaffiliation{NASA Einstein fellow}
\affiliation{Kavli Institute for Cosmological Physics and Enrico Fermi Institute, The University of Chicago, Chicago, IL 60637, USA}

\author{Daniel E. Holz}
\affiliation{Department of Astronomy and Astrophysics, University of Chicago, 5640 South Ellis Ave., Chicago, IL 60637}
\affiliation{Kavli Institute for Cosmological Physics and Enrico Fermi Institute, The University of Chicago, Chicago, IL 60637, USA}
\affiliation{Department of Physics, The University of Chicago, Chicago, IL 60637, USA}

\begin{abstract}
Strong gravitational lensing of gravitational wave sources offers a novel probe of both the lens galaxy and the binary source population. In particular, the strong lensing event rate and the time delay distribution of multiply-imaged gravitational-wave binary coalescence events can be used to constrain the mass distribution of the lenses as well as the intrinsic properties of the source population. We calculate the strong lensing event rate for a range of second (2G) and third generation (3G) detectors, including Advanced LIGO/Virgo, A+, Einstein Telescope (ET), and Cosmic Explorer (CE). For 3G detectors, we find that {$\sim0.1\%$} of observed events are expected to be strongly lensed. We predict detections of {$\sim 1$} lensing pair per year with A+, and {$\sim 50$} pairs {per year} with ET/CE. These rates are highly sensitive to the characteristic galaxy velocity dispersion, $\sigma_*$, implying that observations of the rates will be a sensitive probe of lens properties. We explore using the time delay distribution between multiply-imaged gravitational-wave sources to constrain properties of the lenses. We find that 3G detectors would constrain $\sigma_*$ to {$\sim21\%$ after 5 years}. Finally, we show that the presence or absence of strong lensing {within the detected population} provides useful insights into the source redshift and mass distribution out to redshifts beyond the peak of the star formation rate, which can be used to constrain formation channels and their relation to the star formation rate and delay time distributions for these systems.
\end{abstract}


\section{Introduction}

Strong gravitational lensing is a fundamental measurable property of the Universe. Lensing observables include the fraction of sources which are multiply imaged, as well as statistical distributions of lensing properties such as the image separations and time delays. These are related to the values of the cosmological parameters, as well as the distribution and properties of the matter inhomogeneities which constitute the lenses, ranging from MACHOs (MAssive Compact Halo Objects) and stars to clusters of galaxies. By observing strong lensing, one is able to probe the evolution of the Universe and all matter within it, as well as test the predictions of general relativity.

Observational samples of lensed systems also depend on properties of the sources, and in particular, the number density (for continuous sources such as quasars) or the rate density (for transient sources such as Type Ia supernovae) of the sources as a function of mass and redshift. These samples are also sensitive to observational selection effects, which can cause dramatic differences between the observed and intrinsic lensing distributions.

In the electromagnetic (EM) band, strong gravitational lensing is not only widely used in probing cosmological parameters \citep{1984ApJ...284....1T, 2012JCAP...03..016C, 2020arXiv200808378L}, 
but also in understanding the nature of dark matter halos \citep{2003MNRAS.344.1029D, 2003MNRAS.346..746C, 2003ApJ...599L..61C, 2017ApJS..229...20S, 1994MNRAS.270..245S, 1996MNRAS.283..837S, 2001ApJ...549L..25K, 2002ApJ...568..488O, 2004ApJ...606...67H,2007MNRAS.380..149C, 2010RPPh...73h6901M, 2015ApJ...811...20C, 2018ApJ...857...25D, 2020Sci...369.1347M}. One of {the most basic properties that one can probe are the masses of the lensing halos, as traced by their velocity dispersions, $\sigma$.} 
For example, \citet{2003MNRAS.344.1029D} 
studied 13 lenses provided by 
the Cosmic Lens All-Sky Survey/Jodrell Very Large Array Astrometric Survey data 
to constrain the characteristic velocity dispersion distribution of elliptical galaxies, $\sigma_*$, to  
$\rm 168 \leq \sigma_* \leq 200\ km s^{-1}$ at 68 $\%$ confidence level. 
Similarly, \citet{2005ApJ...630..764C} {selected} $\sim$15 multiply-imaged systems from the same surveys and studied the distribution of the angular separation of these lensing images. By fixing the shape of the galaxy velocity dispersion function either using the Sloan Digital Sky Survey (SDSS) or the Second Southern Sky Redshift Survey (SSRS2), \citet{2005ApJ...630..764C} constrained $\sigma_*$ to $\sim$80 \mbox{km/s} for the case of SDSS, and $\sim$190 \mbox{km/s} for the case of SSRS2. 
In addition, the time delay between lensed images can be used to investigate the density profile of the lens halos as well as the Hubble parameter, $H_0$ \citep{2002ApJ...568..488O, 2012JCAP...11..015L}. 
{Weak lensing surveys, e.g. \citep{To:2020bhf}, provide a complementary probe of the matter distribution at larger scales.}
We note that the distribution of strong lensing of supernovae offers an additional powerful probe~\citep{2001ApJ...556L..71H}, but complete and uniformly selected samples of lensed supernovae continue to pose a challenge. This may change with upcoming surveys, such as those from the Vera Rubin Observatory and {\it Euclid}.

Like electromagnetic waves, 
gravitational waves (GWs) can also be strongly lensed and form multiple images.   
These images appear as separate GW sources with consistent sky positions and binary parameters such as total mass and mass ratio, but with different magnifications and arrival times. The waveforms of multiply-imaged GW sources may also show different phase shifts depending on whether the image is at the minimum, saddle point or maximum of the Fermat potential \citep{Schneider:1992,2020arXiv200712709D}. The magnification changes the overall amplitude of the signal, biasing the inference of the luminosity distance and, as a consequence, the source-frame masses. The time delay affects the arrival time of the lensed signal. Lastly, the phase shift associated to saddle-point images could introduce waveform distortions for signals with higher modes, precession or eccentricity \citep{Dai:2017huk,Ezquiaga:2020gdt}, leading to waveforms which appear to violate general relativity~\citep{Ezquiaga:2020gdt}. All these properties can be used to identify multiple GW events as strongly lensed images of the same source. 

{Strong lensing of GWs} 
will provide a novel and independent way to study the matter distribution in the Universe. 
One advantage over EM studies is that GWs do not suffer from dust extinction or anything else that might compromise the signal; GWs propagate directly from source to observer without any intervening impact (except for the curvature of space-time). 
The correction of dust attenuation in EM observation is a challenging and non-trivial task due to the uncertainty in dust physics {\citep{1997AJ....113..162C, 2000ApJ...533..682C}}. 
Comparing to EM surveys, where it is difficult to guarantee both uniform depth and breadth even for surveys in the radio band {\citep{2019NatAs...3..188A}}, GW detections ``hear'' lensing events happening on the entire sky simultaneously, allowing us to study a clean lensing sample with well-understood and characterized selection effects. Furthermore, unlike EM sources which can be obscured or time variable, the noise power spectrum of GW detectors can be measured and the source properties are well characterized, further reducing selection effects on the lensing sample.
Strong lensing of GW events are sensitive to a wide range of lensing masses, ranging from stellar mass black holes (BHs) to galaxy clusters \citep{2003ApJ...595.1039T, 2012JCAP...11..015L, 2018MNRAS.475.3823S}, 
and will provide important constraints on the underlying dark matter halo distribution in the Universe. {In this paper, we focus on lenses at the scale of massive elliptical galaxies, since these are expected to be the dominant strong lenses. For these systems, the Schwarzschild radius is significantly larger than the wavelength of the GWs emitted by stellar-mass binary black holes (BBHs), and we can therefore adopt the geometric optics limit.}

As mentioned above, EM surveys can use the angular separation between images to constrain the lens population \citep{2003MNRAS.344.1029D, 2005ApJ...630..764C}. 
However, this method does not work for GW detectors due to the large localization errors \citep{Aasi:2013wya}. 
On the contrary, GW facilities have exquisite time resolution (to fractions of a second) which is difficult to achieve in EM surveys even with time variable sources such as quasars or supernovae. We note that the angular separation is proportional to $\sigma^2$ where $\sigma$ is the velocity dispersion of the lens galaxies, while the time delay is proportional to $\sigma^4$. Therefore, time delay distributions are potentially more sensitive to the lens population than angular separation distributions. In what follows we use the time delay distribution between strongly lensed GW events as one of the primary lensing observables.

{A fundamental aspect of statistical lensing is} the rate of strong lensing, which depends both on the properties of the lenses and sources.
Several studies have provided theoretical predictions for this rate. For present 2nd-generation (2G) advanced LIGO (aLIGO), the strong lensing event rate was found to be up to 0.5--1$\,\rm yr^{-1}$ (\citet{2018MNRAS.480.3842O, 2018MNRAS.476.2220L, 2021arXiv210507011Y}). These results are consistent with the current non-detection of lensing events {during the first two observing runs \citep{2019ApJ...874L...2H, 2020PhRvD.102h4031M, 2020arXiv201012093K} and the first half of the third one \citep{Abbott:2021iab}}.\footnote{\citet{2020arXiv200712709D} and \citet{2020arXiv200906539L} have found an intriguing pair, GW170104--GW170814, with masses, sky positions, and phases a priori consistent with the strong lensing hypothesis. However, other properties of the pair such as the large time delay and image type configuration make this association unlikely \citep{2020arXiv200712709D,2020arXiv200906539L}. The analysis of \citet{Abbott:2021iab} confirms that the inclusion of selection effect and source and lens population priors drastically reduce the likelihood that this is a lensing event.} 
The chances of strong lensing will increase with future sensitivity upgrades, as a higher redshift implies a larger probability of lensing. 
2G detectors are expected to be upgraded beyond design sensitivity (A+), which will allow the detection of GW source out to redshift of $z\sim3$ \citep[see Fig.~3 of][]{2019arXiv190403187Tfei}.
Future 3rd-generation (3G) instruments, such as Einstein Telescope (ET) and Cosmic Explorer (CE), 
will be able to detect BBH sources with masses up to $10^4 M_{\odot}$ and at redshifts as high as $z\sim100$ \citep[see Fig.~2 left panel in][]{2020JCAP...03..050M}. 
The enhancement in the detectable cosmological volume will greatly increase the lensing event rate, to as high as 40--$\rm 10^3\,\yr^{-1}$ for ET \citep{2013JCAP...10..022P, 2014JCAP...10..080B, 2015JCAP...12..006D, 2018MNRAS.480.3842O, 2018MNRAS.476.2220L}.

In this work, we explore the capabilities of current and future GW detectors to constrain both the properties of the lens galaxies and the source population. We first compute the lensing optical depth, and calculate the lensing event rates for aLIGO, A+, ET, and CE. We further perform Monte Carlo (MC) sampling to simulate the gravitational lensing of BBHs and calculate the lensing properties including the time delay and magnification distributions. We then estimate our ability to constrain the typical lens velocity dispersion assuming different observation duration times and detector sensitivities. 
Furthermore, since the strong lensing event rate of GWs is also affected by the number of sources in the Universe, we show that this information can be used as a complementary probe of the population of BBH mergers. Both detection and non-detection of GW lensing events will provide insights on the formation channels of these binaries as well as the star formation rate (SFR) and delay-time distributions.\footnote{It is important to note the distinction between the time-delay distribution and the delay-time distribution. The former refers to the amount of time between multiple images of a given strongly-lensed source, {designated by $\delta t$}. The latter refers to the amount of time which elapses between the formation of a binary black hole and the merger of the system, {designated by $\Delta t$.}} 

The paper is organized as follows. In Section \ref{sec:Methods} we
present the methods to calculate the lensing optical
depth, lensing event rate, and lensing simulation, describing in detail our assumptions for both the lens and source population. 
In Section \ref{Sec:results} we show the results for the time delay distributions and lensing rates, discussing their implications to constrain the properties of the lenses and BBH merger sources. We conclude the main results and future prospects in Section \ref{sec:discussion}. 
We adopt the {\it Planck\/} values for the cosmological parameters~\citep{2020A&A...641A...6P}.

\section{Methods}
\label{sec:Methods}

The gravitational lensing of GWs depends both on the population of sources and lenses. In this section we describe the methodology to compute the rate of lensed signals and their properties. We begin in Section \ref{subsec:tau} with computing the probability of strong lensing as determined by the optical depth $\tau(z)$. In Section \ref{subsec: BBHmerger} we provide a prescription for the rate of the BBH merger which acts as GW sources. Fixing the lens model and the source population, we describe the simulation of lensed signals in Section \ref{subsec: sim}. Finally, in Section \ref{sec:lensing_rates} we 
compute the expected strong lensing event rates taking into account the effect of lensing magnification.

\subsection{Lensing optical depth}
\label{subsec:tau}

The probability of a given source at $z_{\rm s}$ being strongly lensed and generating multiple images is determined by the optical depth $\tau(z_{\rm s})$ \citep[see e.g.][]{Schneider:1992}\footnote{It is to be noted that in the limit where the cross-sections significantly overlap with each other $\hat{\sigma}_\text{multiple}$, when $\tau>1$, the probability of lensing is given by $P(z_s) = 1-\exp(-\tau (z_s))$ \citep{Cusin:2019eyv}.}. 
For a given lens model described by a set of parameters $X$, $\tau(z_{\rm s})$ depends on the multiple-image cross section $\hat{\sigma}_\text{multiple}(z_{\rm s},z_{\rm L},X)$ and the density of lenses $n(z_{\rm L},X)$ with properties $X$ at the lens redshift $z_{\rm L}$. The lens density at redshift $z_{\rm L}$ is simply $\int n(z_{\rm L},X) dX$. 
The optical depth is computed directly by adding-up the cross-sections weighted by the density at different redshifts, i.e.
\begin{equation} 
\label{eq:tau_definition}
    \tau(z_{\rm s})=\int_0^{z_{\rm s}}\int \frac{dV_c}{\delta\Omega dz_{\rm L}}n(z_{\rm L},X)\hat{\sigma}_\text{multiple}(z_{\rm s},z_{\rm L},X)\ dX dz_L
\end{equation}

\noindent where $dV_c/(\delta\Omega dz_{\rm L}) = c(1+z)^2D_{\rm L}^2/H(z)$ where $D_{\rm L}$ the angular diameter distance to the lens and $H(z)$ is the Hubble parameter.

In this paper, we choose the singular isothermal ellipsoids (SIE)  \citep{1994A&A...284..285K, 1996astro.ph..6001N, book:411127} as our lens model whose lensing cross-section is determined by their velocity dispersion $\sigma$ and axis ratio $q_g$ of the galaxy. The singular isothermal sphere (SIS) model corresponds to the limit $q_g\to1$. We neglect the shear field since we are less interested in the anisotropic distortion of the signal. 
Qualitatively speaking, the SIE model defines three distinct regions in terms of the number of lensing images in order of increasing area \citep{1994A&A...284..285K}: (1) within the caustic area $\hat{\sigma}_{\rm caustic}$ 4 images form, (2) within the cut region $\hat{\sigma}_{\rm cut}$ 2 images form and (3) in any other region only 1 image forms. 
Therefore, we set $\hat{\sigma}_\text{multiple}=\hat{\sigma}_{\rm cut}$.

The number density of the lens galaxies at redshift $z$ having $\sigma$ and $q_g$ can be described by:
\begin{equation}
\begin{aligned}
n(z_L, X=(\sigma, q_g)) = \phi (\sigma|z_L) p(q_g|\sigma) \end{aligned}
\end{equation}
where $\phi (\sigma|z_L)$ is the number density of the galaxies at a given interval of $\sigma$ at $z_L$, and $p(q_g|\sigma)$ is the distribution of the lens axis ratio for a given $\sigma$. We model $\phi (\sigma|z_L)$, with a Schechter function \citep{1974ApJ...187..425P}:
\begin{equation}
\begin{aligned}
\phi (\sigma|z_L) =  \phi_{*}(z_L)\left(\frac{\sigma}{\sigma_{*}}\right)^{\alpha_{\rm g}} e^{-\left(\frac{\sigma}{\sigma_{*}}\right)^{\beta_{\rm g}}}\frac{\beta_{\rm g}}{\Gamma(\alpha_{\rm g}/\beta_{\rm g})}\frac{1}{\sigma}
\label{eq:Collettpdf}
\end{aligned}
\end{equation}
\noindent where $\phi_*(z_L)$ is the number density of galaxies at redshift $z_L$. In this work, we will consider the case in which the density of galaxies is constant, $\phi_*=8\times 10^{-3} h^3 \Mpc^{-3}$ as measured by \citet{2007ApJ...658..884C}, but our methodology could be extended to include redshift dependence. The power-law index $\alpha_{\rm g}$ and $\beta_{\rm g}$ describe the shape of the distribution \citep{1976ApJ...204..668F, 1977A&A....54..661T}. We set $\alpha_{\rm g} = 2.32$ and $\beta_{\rm g} = 2.67$ also according to the measurement of \citet{2007ApJ...658..884C}.

For a given $\sigma$, the distribution of the lens axis ratio $p(q_g|\sigma)$ which tells the ellipticity of the lens galaxies can be described by a Rayleigh distribution \citep{2015ApJ...811...20C, 2018arXiv180707062H}:
\begin{equation}\label{eq:p_qg}
\begin{split}
p(q_g|s=A+B\sigma) = \frac{1-q_g}{s^2} \exp\left[\frac{-(1-q_g)^2}{2s^2}\right]
\end{split}
\end{equation}

\noindent where $A = 0.38$, $B = {-5.7 \times 10^{-4} (\mbox{km/s})^{-1}}$ \citep{2015ApJ...811...20C}{, implying that more massive galaxies are more spherical.} We set the minimum $q_{\rm g, min}=0.2$. 

The angular scale of the lensing cross-section is determined by the angular Einstein radius:  
\begin{equation}
\begin{aligned}
\theta_E = 4\pi \left( \frac{\sigma}{c} \right)^2 \frac{D_{\rm LS}}{\rm D_S}\,,
\label{eq:theta E}
\end{aligned}
\end{equation}
where $D_{\rm LS}$ is the angular diameter distance between the lens and the source, and $D_{\rm S}$ is the angular diameter distance between the observer and the source. 
Apart from the geometrical configuration of the source-lens system, the Einstein radius is fully determined by the galaxy velocity dispersion, $\sigma$. 
This scale is the same for both SIS and SIE. 
The multiply-lensed cross-section for SIE is then given by:
\begin{equation}
    \hat{\sigma}_\text{multiple}(z_{\rm s},z_{\rm L},\sigma,q_g)={ \theta^2_E(z_{\rm s},z_{\rm L}, \sigma)\tilde{\sigma}_{\rm cut}(q_g)},\
\end{equation}
where $\tilde{\sigma}_{\rm cut}(q_g)$ is the dimensionless cut cross-section given by \citep{1994A&A...284..285K} in units of $\theta_E$:
\begin{equation}
\begin{aligned}
    \tilde{\sigma}_{\rm cut}(q_{g})\ = \frac{4q_{g}}{1-q_{g}^2}\int_{q_{g}}^1 \frac{\rm arccos \Delta }{\sqrt{\Delta^2-q_{g}^2}} d\Delta.
\end{aligned}
\end{equation}
\vspace{1pt}

\noindent This quantity depends only on $q_g$. {Note that in the limit of a spherical lens, $q_g\to1$, we find that $\tilde{\sigma}_{\rm cut}\to\pi$ and we recover the usual SIS cross-section}. The SIS model has two regions delimited by the Einstein radius, where 2 images form inside and 1 outside; its cross-section does not depend on $q_g$.

Combining all of the ingredients above, we now define the optical depth for multiple-images:
\begin{widetext}
\begin{equation}
\begin{split}
\tau(z_{\rm s}) = \int_{0}^{z_{\rm s}}&\int_{\sigma_{\rm min}}^{\sigma_{\rm max}}\int_{0.2}^1  16\pi^3\frac{c(1+z_L)^2}{H(z_L)}\left(\frac{D_LD_{LS}}{D_S}\right)^2\left(\frac{\sigma}{c}\right)^4\phi(\sigma|z_L) p(q_g|\sigma)\tilde{\sigma}_{\rm cut}(q_g)dq_g d\sigma dz_{\rm L}\,,   
\label{eq:tau1}
\end{split}
\end{equation}
\end{widetext}
\noindent which integrates all the cross-sections of the lens galaxies between the observer and the source.

Note that in the SIS limit, the dependence on $q_g$ disappears and one can get a closed form result by integrating in terms of Gamma functions. 
This result is subject to $\sigma_{\rm max}$ and $\sigma_{\rm min}$, the upper and lower bounds of the velocity dispersion of the lens galaxies. For simplicity, we fix $\sigma_{\rm max} = \infty$ and $\sigma_{\rm min} = 0$. However, other values are possible. For example, $\sigma_{\rm min} \sim 70 \rm \mbox{km/s}$ might be more consistent with observations \citep{2007ApJ...658..884C, 2013ApJ...764..184M}. 
We discuss the effect of changing $\sigma_{\rm max}$ and $\sigma_{\rm min}$ on $\tau(z_s)$ in Appendix \ref{app:tau}.

\begin{figure}[t!]
	\includegraphics[width=1.\columnwidth]{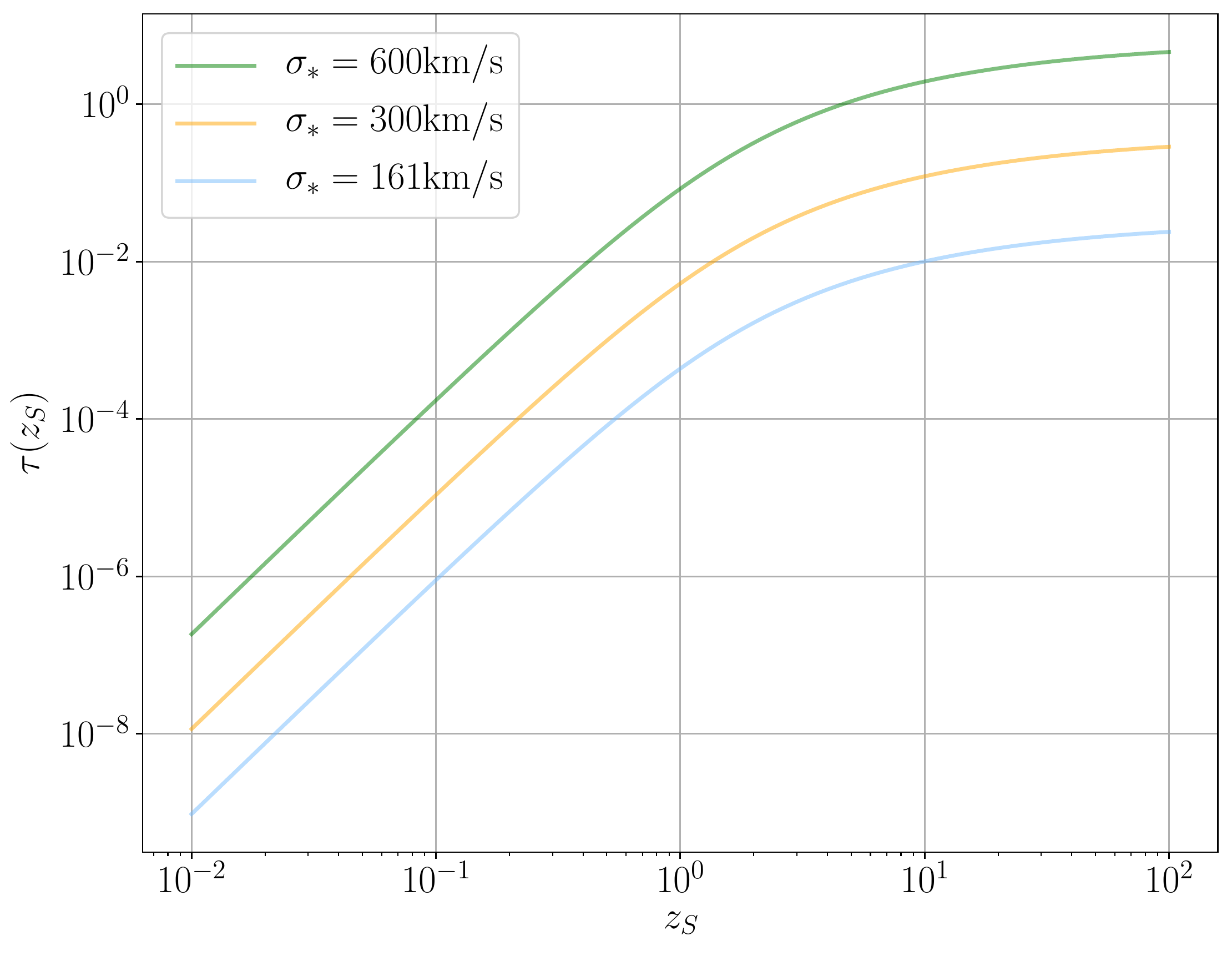}
    \caption{Optical depth $\tau$ as a function of source redshift $z_s$ with different $\sigma_{*}$ represented by different colors. Increasing $\sigma_{*}$ will increase the velocity dispersion of the whole galaxy population, hence increase the lensing cross-sections.}
    \label{fig:optical_depth}
\end{figure}

Adding all the pieces together, Figure \ref{fig:optical_depth} shows the optical depth $\tau(z_s)$ assuming 3 different values of $\sigma_*$. In general, $\tau(z_s)$ increases with $z_s$ because there are more intervening galaxies between the source and the observer at higher $z_s$. At a given $z_s$, $\tau(z_s)$ increases with increasing $\sigma_*$ since the lensing cross-section of the galaxy population increases with $\sigma_*$.
We also find that $\tau(z_s)$ can be well-approximated by the optical depth of the SIS model ($\tau_{\rm SIS}$) multiplied by a constant factor {$\sim0.96$}. We elaborate more on these differences in the optical depth between SIS and SIE lens model in Appendix \ref{app:tau}.

\subsection{Source population: binary black holes}
\label{subsec: BBHmerger}

Once we know how to compute the probability of strong lensing, the next ingredient is to model the population of sources. This information will be later used to simulate lensed events and to compute the lensing rates and distributions. 
We begin with the differential merger rate as a function of observing time $t$ (in detectors frame) is given by
\begin{equation}
 \frac{d\dot{N}(z)}{dz}\equiv\frac{d^2N}{dzdt}(z) =\frac{\mathcal{R}(z)}{1+z} \frac{dV(z)}{dz}
\end{equation}
\noindent where $\mathcal{R}(z)$ describes the source-frame merger rate density, $\frac{dV}{dz} = 4 \pi c \frac{r_c^{2}(z)} {H(z)}$ is the differential comoving volume, and the $(1+z)$ factor converts from source frame to detector frame. In this work, we fix the local merger rate density $\mathcal{R}(z=0) \equiv \mathcal{R}_0 = {64.9}_{-33.6}^{+75.5} \Gpc^{-3} \yr^{-1}$ \citep{2019ApJ...882L..24A}.

In order to model the redshift evolution of the merger rate we will follow two complementary approaches. First, we will consider a model in which the BBHs are assumed to follow the SFR with an additional delay time. 
The delay time $\Delta t$ is the time between the binary formation and the final merger. 
This is motivated by the assumption that BBHs are formed from stars in the field and has been studied thoroughly using population synthesis codes \citep{2002ApJ...572..407B, 2014LRR....17....3P}. 
{Observations of strongly lensed events will provide constraints on both the SFR and the delay-time. If one believes we already know the SFR, then our results probe the delay-time distribution directly. {These constraints would be complementary to the ones obtained with unlensed, low-redshift binaries \citep{Fishbach:2021mhp}.} 
Alternatively, prior knowledge of the delay-time distribution would allow for direct constraints on the SFR of the sources.
In our analysis we consider three different scenarios for the SFR and delay-time distribution, to explore the impact that these have on our results.} 
In the main text of the paper, we adopt the SFR model from \citet{2014ARA&A..52..415M} (MD14) with minimal delay time $\Delta t_{\rm min}$= 50 Myr. We discuss two additional scenarios in the Appendix \ref{app:Rz}: MD14 SFR model with a different delay time of $\Delta t_{\rm min}$= 1 Gyr, and a different SFR density which is constant throughout the redshift evolution $\dot{\rho_{*}} = 0.004\,M_{\odot}\Mpc^{-3} \yr^{-1}$ with $\Delta t_{\rm min}$= 50 Myr. The detailed calculation of the rate from the SFR to detector-frame merger rate is described in Appendix \ref{app:Rz}.

Our second approach will be to extend this fixed model by varying its elements in a convenient parametrization from \citet{2020ApJ...896L..32C}: 
\begin{equation}
\mathcal{R}(z | z_p, \alpha, \beta) = \mathcal{C}(\alpha, \beta, z_p)\frac{\mathcal{R}_0 (1+z)^{\alpha}}{1+(\frac{1+z}{1+z_p})^{\alpha+\beta}}
\label{eq:RBBH}
\end{equation}
\noindent where $\mathcal{C}(\alpha, \beta, z_p) = 1+(1+z_p)^{-\alpha-\beta}$. Equation \ref{eq:RBBH} is proportional to $(1+z)^{\alpha}$ at low redshift and $(1+z)^{\beta}$ at high redshift. $z_p$ is the redshift at the peak of the distribution, and the local merger rate $\mathcal{R}(z=0) =\mathcal{R}_0={64.9}_{-33.6}^{+75.5} \Gpc^{-3} \yr^{-1}$ is fixed \citep{2019ApJ...882L..24A}. The second approach will be relevant when quantifying how the source population affects the lensing rate, as discussed in Section \ref{sec: res souce_pop_Rlens}. {We note that alternate formation channels might be described with differing values of $\alpha$, $\beta$, $z_{p}$, and $\mathcal{R}_0$, or with entirely different functional forms. These could be combined to generalize our approach; for this paper we describe the aggregate population with a single distribution shown in Equation ~\ref{eq:RBBH}.}

In order to calculate how many of these BBH mergers are detected, we need to consider the detection probability, $p_\text{det}(\mathcal{M}, q, z)$, which takes into account the detector sensitivity and selection bias for binaries with different masses and redshifts. We parametrize the source masses in terms of the chirp mass, $\mathcal{M}=(m_1m_2)^{3/5}/(m_1+m_2)^{1/5}$, and mass ratio, $q=m_2/m_1$, where $m_1$ is the mass of the heavier BH, $m_1>m_2$. The detected merger event rate within redshift $z$ is given by:
\begin{widetext}
\begin{equation}
\begin{aligned}
\dot{N}_\text{BBH}(z) = \int_0^z\int_{\mathcal{M}_{min}}^{\mathcal{M}_{max}}\int_{0}^{1}\frac{d\dot{N}(z)}{dz}p(\mathcal{M}, q)p_\text{det}(\mathcal{M}, q, z)\,dq\,d\mathcal{M}\,dz \,,
\label{eq:NBBH}
\end{aligned}
\end{equation}
\end{widetext}
where $p(\mathcal{M}, q)$ is the 2-dimensional distribution of $\mathcal{M}$ and $q$.
We assume $m_1$ follows a power-law distribution $p(m_1) \propto m_{1}^{-0.4}$ and $m_2$ is uniformly distributed in range $m_{\rm min}<m_2<m_{1}$. We fix $m_{\rm min} = 5 M_{\odot}$ and $m_{\rm max} = 41.6 M_{\odot}$ 
following the results of the first and the second observing run of Advanced LIGO and Advanced Virgo \citep{2019ApJ...882L..24A}. We derive the distribution of $\mathcal{M}$ and $q$ by randomly drawing $m_1$ and $m_2$ and then linearly interpolate the 2-dimensional probability density function (PDF) to get $p(\mathcal{M}, q)$, and also the corresponding minimum and maximum $\mathcal{M}$, $\mathcal{M}_{\rm min}$ and $\mathcal{M}_{\rm max}$. 
We note that the latest LIGO--Virgo catalog, GWTC-2, provides a more complex description of the mass distribution \citep{Abbott:2020gyp} and in fact this simple power-law model is disfavored by observations. 
However, for the purposes of our analysis this simplified description is sufficient. 

We determine the probability of detecting a given source by the fraction of events across all possible sky-locations, orientations, and inclinations that are above a given signal to noise threshold $\rho_\text{\rm thr}$. For a particular detector/detector network this is a known function \citep{2015ApJ...806..263D}:
\begin{equation}
    p_\text{det}(\mathcal{M}, q, z) = P(w=\rho_\text{thr}/ \rho_\text{opt}(\mathcal{M}, q, z))\,,
\label{eq: Pdet}
\end{equation}

\noindent where $\rho_{\rm opt}(\mathcal{M}, q, z)$ is the S/N for an optimally located and oriented binary. 
We focus on a single detector with threshold of $\rho_{\rm thr}=8$ and consider 4 sensitivies: aLIGO \citep{2015CQGra..32g4001L}, aLIGO at upgraded sensitivity (A+) \citep{2019arXiv190403187T} and the third generation detector Einstein Telescope (ET) \citep{2020JCAP...03..050M} and Cosmic Explorer (CE) \citep{2019BAAS...51g..35R}.\footnote{The sensitivity curve ($S_{h}(\text{Hz}^{-1/2})$) for different detectors are from: https://dcc.ligo.org/LIGO-T1500293-v11/public} {We do not take into account the duty cycle and assume that the detectors are always online.} 

\subsection{Simulating strongly lensed GW events}
\label{subsec: sim}

Having fixed the lens model (SIE model) and the source population (BBHs consistent with LIGO/Virgo O2), we now describe our method for generating the sample of strong lensing events. We adopt a semi-analytical approach similar to that in \citet{2018arXiv180707062H} which randomly generates lens systems and solves the corresponding lens equations. The detailed procedure of the MC simulation can be found in Section 2 of Appendix A in \citet{2018arXiv180707062H}. We highlight the differences in our simulation below: 
\begin{enumerate}
    \item We sample the BBH mass $m_1$ and $m_2$ using the distribution described in Section \ref{subsec: BBHmerger}.
    
    \item We pick the source redshift ($z_s$) based on the PDF normalized from the BBH merger rate as a function of redshift $\dot{N}_{\rm BBH}(z)$ calculated in Section \ref{subsec: BBHmerger}.
    
    \item Since we want to constrain lens parameter $\sigma_*$, we directly pick velocity dispersions of the galaxy lenses based on the PDF normalized from the Schechter function in Equation \ref{eq:Collettpdf} with varying $\sigma_*$ values instead of setting $\rm \sigma_* = 161  \mbox{km/s}$ as in \citet{2018arXiv180707062H}. 
    
    \item {{Our lensing simulation assumes that multiple images of the same source have independent detector selection effects.} {Since lensed images of the same source arrive at different times, the relative angles between the detector and the source will have changed, and thus the detector response will be different for the two images.} {We note, however, that since the images come from the same binary source, the intrinsic angles of the binary source will be the same. It is computationally expensive to incorporate this, and since we do not expect these correlations to qualitatively impact any of our results, we neglect them.} To determine whether a lensing image can be detected or not, we generate one random number for each image respectively. If the random number is smaller than $P(w = \rho_{\rm thr}/\sqrt{\mu}\rho_{\rm opt})$, we consider the image have been detected.} {\citet{2019ApJ...874..139Y} show that incorporating the Earth's rotation decreases the lensing event rate by $\sim 10\%$ for the case of BBHs.}
\end{enumerate}

After picking the parameters for the sources and the lens galaxies, we follow the procedure in \cite{2018arXiv180707062H} and randomly draw $z_{\rm L}$ and pick the source-plane location where we can find multiple images. We obtain the image positions $x_{1,i}$ and $x_{2,i}$ for the $i$-th image ($i$=1, 2 for the case with 2 images, or $i$=1, 2, 3, 4 for case with 4 images) by solving the lens equations (see Equations 11--14 in \citet{2018arXiv180707062H}) and calculate the magnification for each image:

\begin{equation} 
\begin{aligned}
\mu_{i} = \left(1- \sqrt{\frac{q_g}{x_{1,i}^2+q_g^2 x_{2,i}^2}}\right)^{-1}
\label{eq:mu}
\end{aligned}
\end{equation}
\noindent and the time delay for $i$th image relative to a reference time (see more details in \citet{1994A&A...284..285K}):
\begin{equation}
\begin{aligned}
 \delta t_{i} = 16 \pi^2 \frac{D_c(z_{L})}{c} \left(\frac{\sigma}{c} \right)^4 \left( 1-\frac{D_c(z_{L})}{D_c(z_{s})} \right) \Phi_{i}
\label{eq:delt}
\end{aligned}
\end{equation}
\noindent where $D_c(z_s)$ is the comoving distance of the source, $D_c(z_L)$ is the comoving distance of the lens, and $\Phi_i$ is the Fermat potential \citep{1986ApJ...310..568B}.

One of the goals of this work is to explore the ability of GW detectors to constrain the characteristic galaxy velocity $\sigma_{*}$ by observing the time delay ($\delta t$) distribution of multiply-lensed events. 
In particular, we focus on the time delay between {two} detected lensing images from the same source: 
\begin{equation}
\begin{aligned}
\delta t = |\delta t_{1}-\delta t_{2}|\propto \sigma^4\,.
\end{aligned}
\end{equation}
{In most of the cases, these two images correspond to the primary (the brightest image, or the one with the highest magnification) and the secondary image (the second-brightest image) except for some very rare cases.} 
This time delay $\delta t$ should not be confused with the delay time $\Delta t$ between the formation and merger of binary black holes introduced in Section \ref{subsec: BBHmerger}.

\begin{figure}
\centering
	\includegraphics[height = 6.5cm, width=1\columnwidth]{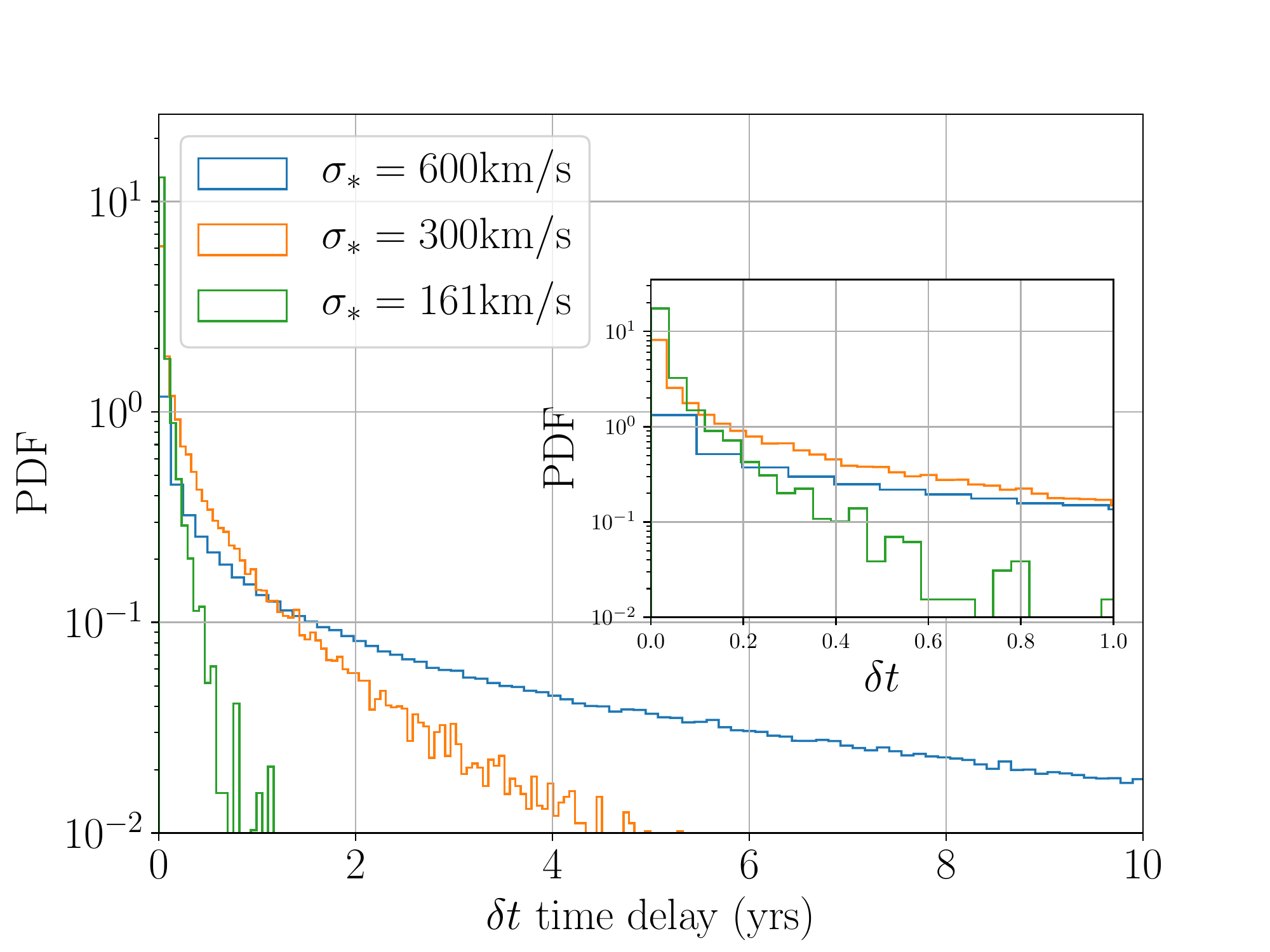}
	\includegraphics[height = 6.5cm, width=1\columnwidth]{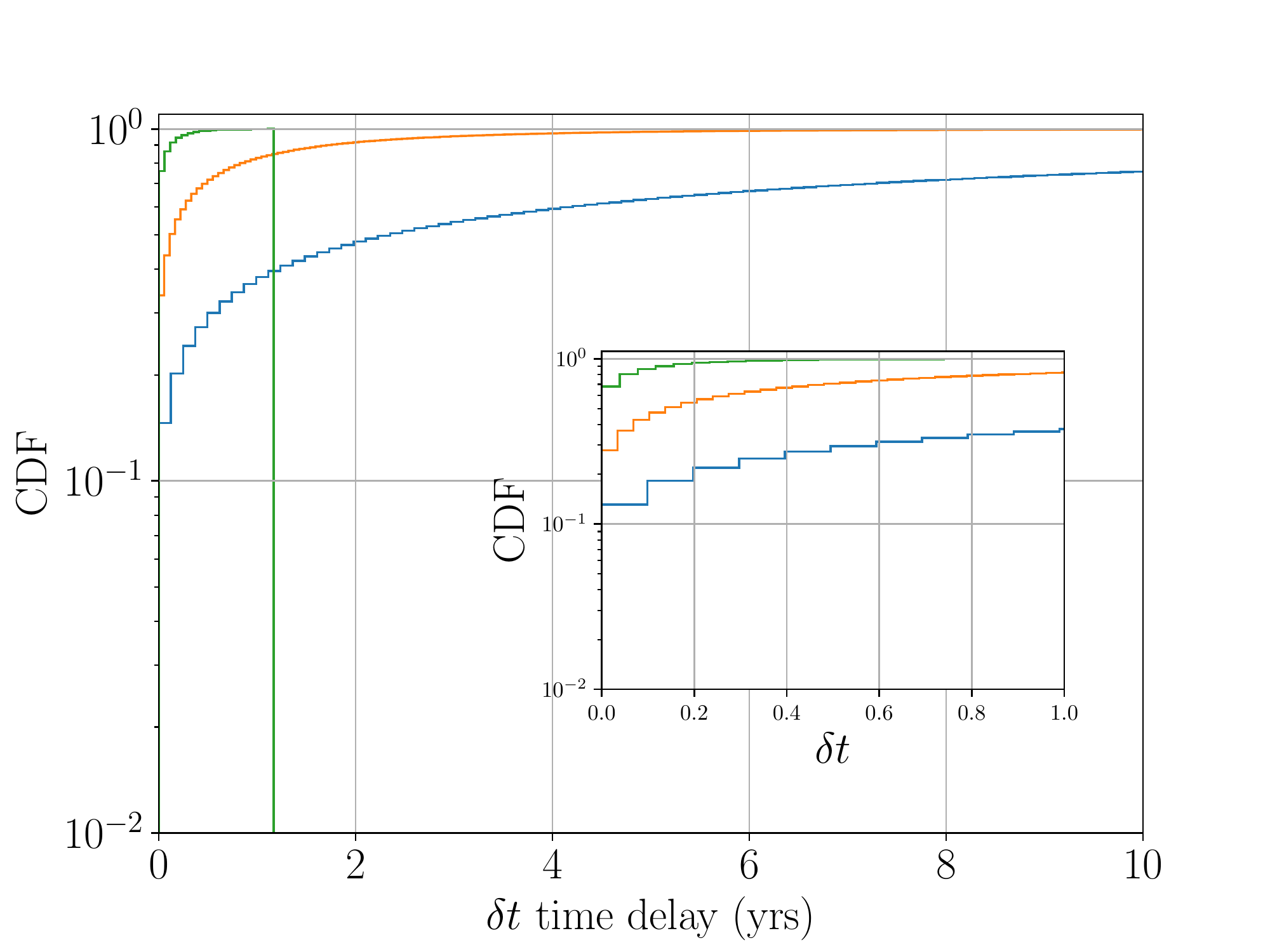}
    \caption{Lensing time delay $\delta t$ distribution for strong lensing pairs observed by ET for $\sigma_{*} = 161, 300, 600$ \mbox{km/s} assuming $10^7$ BBH sources. The top panel displays the PDF, while the bottom one plots the CDF. Time delay extends to higher values when we increase $\sigma_{*}$. The proportionality between time delay and $\sigma$ is described in Equation (\ref{eq:delt}). {The green CDF ($\sigma_{*} = 161 \rm \mbox{km/s}$) truncates at the maximum $\delta t$.} We generate the BBH population using the MD14 SFR model \citep{2014ARA&A..52..415M} assuming merger delay time distribution $P(\Delta t) \propto 1/\Delta t$ ranging from $50$ Myr to 13.5 Gyr (See more details in Section \ref{subsec: BBHmerger}).}
    \label{fig:time_delay_distr}
\end{figure}

Since $\delta t_i$ is proportional to $\sigma^4$ according to Equation \ref{eq:delt}, the time delay distribution is very sensitive to the value of $\sigma_{*}$. By comparing the observed $\delta t$ distribution with the theoretical prediction for different $\sigma_*$, we can then constrain the value of $\sigma_{*}$. 
We present the PDF distribution of $\delta t$ for 3 different $\sigma_*$ in Figure \ref{fig:time_delay_distr}. 
When $\sigma_*$ is high, the $\delta t$ distribution extends to higher values. For the case of $\rm \sigma_* = 600 \mbox{km/s}$ the tail of the $\delta t$ distribution extends to even 15 years. 
To facilitate the visualization, we zoom in to the range $<$ 1 year in the inset of the same figure. In general, higher $\sigma_*$ has a higher probability of high $\delta t$ values. The cumulative distribution function (CDF) also has noticeable differences. The $\delta t$ at which 90 $\%$ of the events are included for $\sigma_* = 161, 300, 600$\mbox{km/s} are 0.16, 1.77, and 25.72 years.

To constrain $\sigma_*$ using GW lensing events, we perform a Kolmogorov-Smirnov (KS) test which is a widely used statistical technique to quantify the difference between the model and the data. 
The KS test computes the distance between the CDF of the model and the empirical probability distribution (EDF) of the data \citep{Kolmogorov, smirnov1948}. The maximum distance is defined as the KS statistic value. Bigger KS statistic indicates that the 2 input distributions may have different origins. If a continuous expression for the model CDF is not available, we can apply the two-sample KS test which uses EDF of the theoretical data set instead of the CDF. Since we do not have an analytical expression for the lensing $\delta t$ distribution and we do not want to add additional uncertainties by fitting the theoretical $\delta t$ distribution from the simulations, we adopt two-sample KS test in the following analysis.

Operationally, we generate mock observation samples and compare them with the theoretical $\delta t$ distribution to get a distribution of KS test statistics. We denote the $\sigma_*$ used for generating the theoretical $\delta t$ distribution as $\sigma_{*,A}$, and for the mock observation distribution as $\sigma_{*, B}$. The KS statistic distribution from comparing the theoretical distribution with $\sigma_{*,A}$ and the mock observation distribution with $\sigma_{*, B}$ is expressed as KS($\sigma_{*, A}$, $\sigma_{*, B}$). We use the KS statistics when $\sigma_{*, A} = \sigma_{*, B}$ (i.e. KS($\sigma_{*, B}$, $\sigma_{*, B}$)) as a reference. If the majority of the KS($\sigma_{*, A}$, $\sigma_{*, B}$) derived from observation samples are greater than the majority of KS($\sigma_{*, B}$, $\sigma_{*, B}$), then it indicates that $\sigma_{*, B}$ is actually quite different from $\sigma_{*, A}$, implying the observed $\sigma_*$ is inconsistent with the theoretical prediction. 

The size of each lensing $\delta t$ distribution sample is determined by the BBH merger rate, the optical depth, the observation duration time, and the detector sensitivity. Due to the low lensing event rate of aLIGO and A+, we only discuss the possibility of using $\delta t$ distribution to constrain galaxy population using 3G detectors. In particular we concentrate on ET as an example, although similar results are expected for CE. The lensing time delay can sometimes be larger than the observation duration time. In order to make the sample realistic, we exclude the sources that have time delay greater than the observation duration time. 

We generate the theoretical $\delta t$ distributions by simulating a large ($10^7$) number of sources. For the mock observation samples, we set the number of sources going into our simulation using the product of the BBH merger rate per year as calculated in Section \ref{subsec: BBHmerger} and the observation duration time ranging from 1 year, 5 years, and 10 years. 

As a summary, we follow the procedure below to test the consistency of the model and the mock sample: 
\begin{enumerate}
    \item For a given observation duration time, we generate 500 mock $\delta t$ distribution samples for a given lens galaxy population with $\sigma_{*,B}$. We exclude the sources that have time delay greater than the observation duration time. 

    \item We compare these mock samples with the theoretical $\delta t$ distribution using the KS test. For a given observation time, $\sigma_{*, A}$, and $\sigma_{*, B}$, we can get 500 KS statistic values and derive their corresponding PDF. 
    
    \item We use the PDF of the KS test values for the case where $\sigma_{*, A} = \sigma_{*, B}$ as the reference distribution. The distribution of KS test statistics shifts to larger values when $\sigma_{*, A} \neq \sigma_{*, B}$. We can also compute the distribution of the ratio of the KS statistics KS($\sigma_{*, A} = \sigma_{*, B}$, $\sigma_{*, B}$)/KS($\sigma_{*, A}$, $\sigma_{*, B}$). Most of the time the ratio should be smaller than 1 because the mock samples are usually closer to the theoretical models with the same value of $\sigma_{*}$. However, sometimes due to the limitation of the observation time, the observation sample may appear closer to the wrong model. We define the area where the PDF of this ratio is smaller than 1 as the probability of correct inference. We show how the probability of correct inference evolves with the observation duration time in Section \ref{sec: res delta t}. 
\end{enumerate}

In addition to the $\delta t$ distribution, another interesting observable is the relative magnification distribution: the ratio of the magnification of the secondary image $\mu_2$ and the primary image $\mu_1$, $\mu_2/\mu_1$. Since it is not directly related with $\sigma_*$ but more sensitive to the ellipticity of the lenses, we discuss them in the Appendix \ref{app:mag ratio}. 
It would be interesting to combine both observables in future analyses to constrain the lens population more comprehensively.

\subsection{Computing strong lensing event rates}
\label{sec:lensing_rates}

In this section, we focus on the calculation of the \emph{observed} GW strong lensing event rate $\dot{N}_\text{lensing}$. To achieve this, we need to take into account how many merging sources are multiply-imaged, as well as which of these sources are detectable. 
We thus include both the optical depth $\tau(z)$ and the magnification distribution $P(\mu)$ into the integration in Equation (\ref{eq:NBBH}): 
\begin{widetext}
\begin{align}
\dot{N}_\text{lensing}(z) = 
\int_0^z\int_{\mathcal{M}_{min}}^{\mathcal{M}_{max}}\int_{0}^{1}\int_{\mu_{min}}^{\mu_{max}}
&\quad \tau(z) \frac{d\dot{N}(z)}{dz}p(\mathcal{M}, q)p_\text{det}(\mu, \mathcal{M}, q, z) P(\mu) d\mu dq  d\mathcal{M} dz
\label{eq:Rlensing_mag}
\end{align}
\end{widetext}
where $p_\text{det}(\mu, \mathcal{M}, q, z)$ is modified due to the magnification factor $\mu$ as follows:
\begin{equation}
    p_\text{det}(\mu, \mathcal{M}, q, z) = P \left(w=\frac{\rho_\text{thr}}{\sqrt \mu \rho_\text{opt}(\mathcal{M}, q, z)} \right)\,,
\end{equation}
where we have changed $\rho_{\rm opt}$ to $\sqrt{\mu}\rho_{\rm opt}$. This is because magnifying a source with factor $\mu$ is equivalent to decreasing the source luminosity distance by a factor of $1/\sqrt{\mu}$, and the luminosity distance enters in the signal-to-noise via $\rho\propto 1/d_L$. 

The values and meaning of the strong lensing event rate depends critically on the choice of the magnification distribution $P(\mu)$. 
For example, \citet{2018MNRAS.480.3842O} uses two differing magnification distribution when calculating $\dot{N}_\text{lensing}$. The first way is treating all the images from the same BBH source as a single group and use the sum of the magnification values as the total magnification. 
Another way is treating individual images differently which means defining $P(\mu)$ using the magnification value of each image regardless of the source. \citet{2017PhRvD..95d4011D} propose a fitting to the magnification PDF at different redshifts based on the simulations in \citet{2008MNRAS.386.1845H} and \citet{2011ApJ...742...15T}. \citet{2018MNRAS.476.2220L} adopts the magnification of the fainter image in the case of double images and that of the third brightest image in the case of four lensing images. Many other works (e.g. \citet{Ng:2017yiu, 2019A&A...625A..84D}, etc) prefer using a simple analytical form of $P(\mu) \propto \mu^{-3}$ to describe the tail of the magnification distribution at high values, typically applied for $\mu>2$. 

In real observations, the identification of strongly lensed GW events is not an easy task. One needs to statistically asses if the parameters of each possible image favor the lensing hypothesis over the non-lensed hypothesis \citep{2019ApJ...874L...2H}. 
This is typically achieved by searching for overlaps in the sky maps, masses, and spins. However, this overlap in binary paramter space can also happen in non-lensed events due to selection effects and observational errors. In addition, one could also identify lensed GW event alone without associating it with other events by measuring the phase distortion with respect to the unlensed predictions in general relativity. However, this is only applicable for type II images which are created at the saddle points of the time delay surface. They have a phase shift of $\pi/2$ which modifies the phase evolution when higher modes, precession or eccentricity are present \citep{2017arXiv170204724D, 2020arXiv200812814M}. 
3G detectors could be sensitive to these distortions, identifying type II images directly \citep{Wang:2021kzt}. 
Although identifying type II images individually could help constraining the optical depth, in order to measure the time delay distribution we need to identify at least two images of the same source. It is important to remember that the first image, typically the brightest one, is always at the minimum of the time delay surface (type I) \citep{Blandford:1986zz} and thus cannot be identified individually.

Considering above, we calculate two kinds of lensing event rates. The first one is the number of lensed systems whose primary images are detected per year, denoted as $\dot{N}_{\rm lensing, 1st}$.  
The second one is the lensing event rate when at least 2 images are detected for each lensing system, $\dot{N}_{\rm lensing, 2nd}$.  The first quantity $\dot{N}_{\rm lensing, 1st}$ is useful to understand how the observed BBH population is "polluted" by magnified events, since the primary images with the largest magnification are the most likely ones that can be detected but may not be identified as lensing events if we miss the other images associated with the same source. To calculate $\dot{N}_{\rm lensing, 1st}$, we use the magnification distribution of the primary image $P(\mu)_{\rm 1st}$ when doing the integration in Equation \ref{eq:Rlensing_mag}. The second quantity $\dot{N}_{\rm lensing, 2nd}$ is useful to know how many multiply-lensed events we can detect for studying $\delta t$ distribution. To estimate this one, we use the magnification distribution of the secondary image $P(\mu)_{\rm 2nd}$. The idea is that if the secondary image can be detected, then the primary image is very likely to be detected as well since by definition the primary image should have a higher $\sqrt{\mu}\rho_{\rm opt}$. {Nevertheless, we notice that due to the difference in arrival times, the orientation angle of the detector will change before the second image arrives. There is a possibility that we detect the secondary image but the first one arrives when the orientation of the detector network is less favorable. It is also likely that the first one is missed because the detector is not online but in this paper we assume the detector observes whole year.
However, we show in Appendix \ref{app:Rate lensing} that this procedure still gives a very good estimation of $\dot{N}_{\rm lensing, 2nd}$ when comparing to our lensing simulations.} 

\begin{figure}
	\includegraphics[height=0.8\columnwidth, width=\columnwidth]{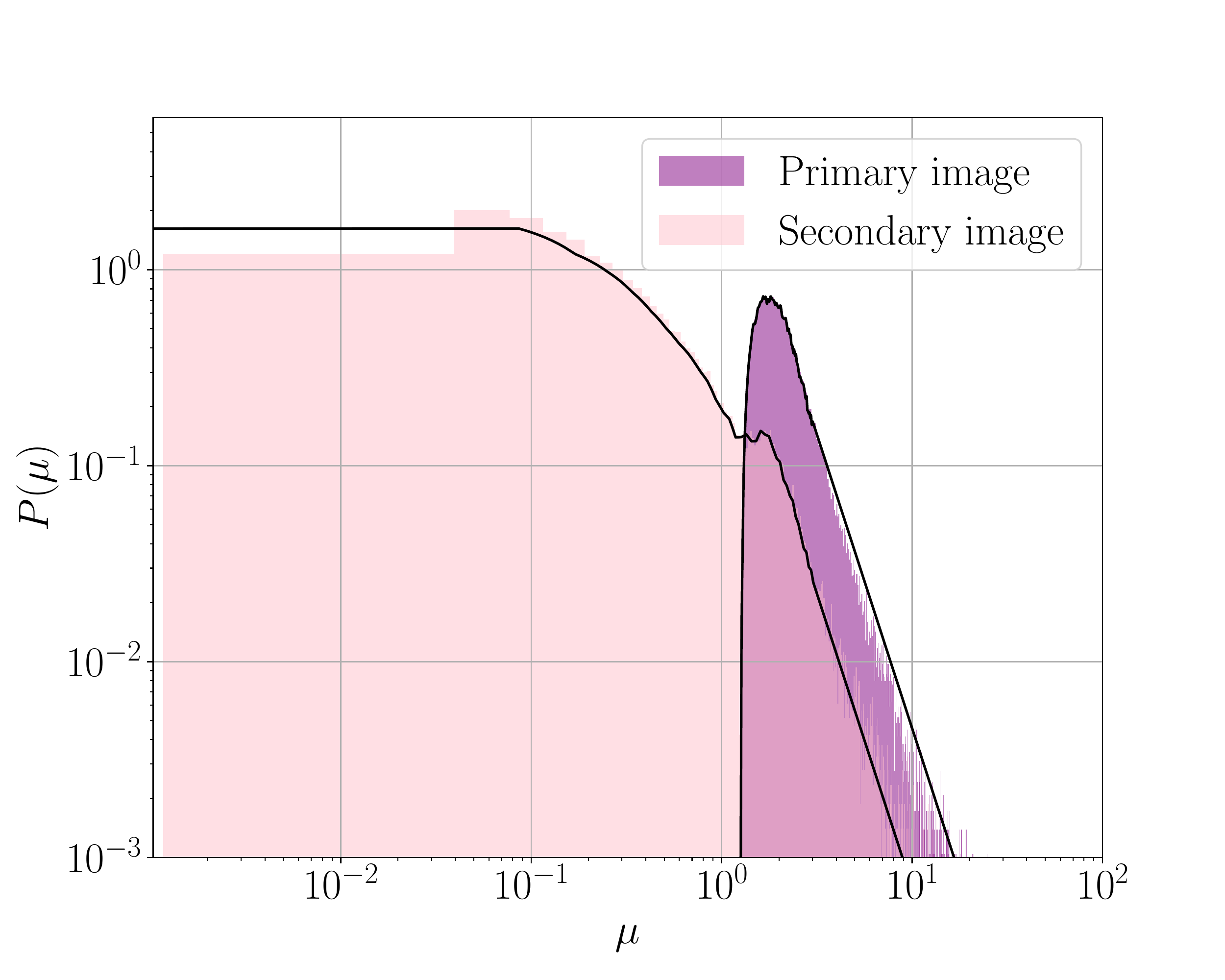}
    \caption{Magnification distribution, $P(\mu)$, obtained from our MC simulations of GWs lensed by SIE lenses. Purple and pink histograms correspond to the primary ($P(\mu)_{\rm 1st}$) and secondary ($P(\mu)_{\rm 2nd}$) images, which correspond to the brightest and second-brightest images respectively. We compute $P(\mu)$ by linearly interpolating the histogram for $\mu <3$ and smoothly connect it with a power-law function $P\propto \mu^{-3}$ for $\mu > 3$. The final $P(\mu)$ we use in Equation \ref{eq:Rlensing_mag} are marked by black solid lines. We set $\mu_{\rm min}$ and $\mu_{\rm max}$ based on the $P(\mu)$ derived from the lensing simulation.} 
    \label{fig:Pmu}
\end{figure}

In practice, we obtain $P(\mu)_{\rm 1st}$ and $P(\mu)_{\rm 2nd}$ from our lensing simulation by recording the magnification of the brightest image and the second-brightest image of each lensing system. We compute the histograms of $\mu_1$ and $\mu_2$ and linearly interpolate them for $\mu < 3$. We smoothly connect them with a power law $\mu^{-3}$ for $\mu > 3$, as it is universally expected for large magnifications. Figure \ref{fig:Pmu} shows the results. 
{It is to be noted that the distribution of secondary images {extends to $\mu_2 <1$}. {This is because the secondary images of SIE model are very close to the lens center and thus are highly de-magnified \citep{1994A&A...284..285K}.} On the other hand, the primary image magnification distribution peaks at $\mu\sim2$.} We summarize our results in Section \ref{sec: Rlensing}.

\section{Results}
\label{Sec:results}

In the previous section we have introduced our procedure for calculating lensing event rates and lensing distributions given a lens and source population and a detector sensitivity. 
In this section we present our results and discuss how the lens and source populations affect the lensing observables, particularly the lensing event rate as a function of redshift and the lensing time delay distribution. 
We also examine the capacity of present and future detectors to probe these lens and source parameters.

We show the calculation of the lensing rates and its dependence on lens velocity dispersion parametrized by $\sigma_*$ in Section \ref{sec: Rlensing}. In Section \ref{sec: detectable cosmo volume} we show how the detectable cosmological volume is expanded by the detection of lensing events. Section \ref{sec: res delta t} demonstrates the potential constraints on $\sigma_*$ from the time delay distribution. Lastly, we discuss how the lensing event rate is affected by variations in the source population in Section~\ref{sec: res souce_pop_Rlens}.

\subsection{Lensing event rate}
\label{sec: Rlensing}

\begin{figure*}
\centering
\includegraphics[height=9cm,width=11cm]{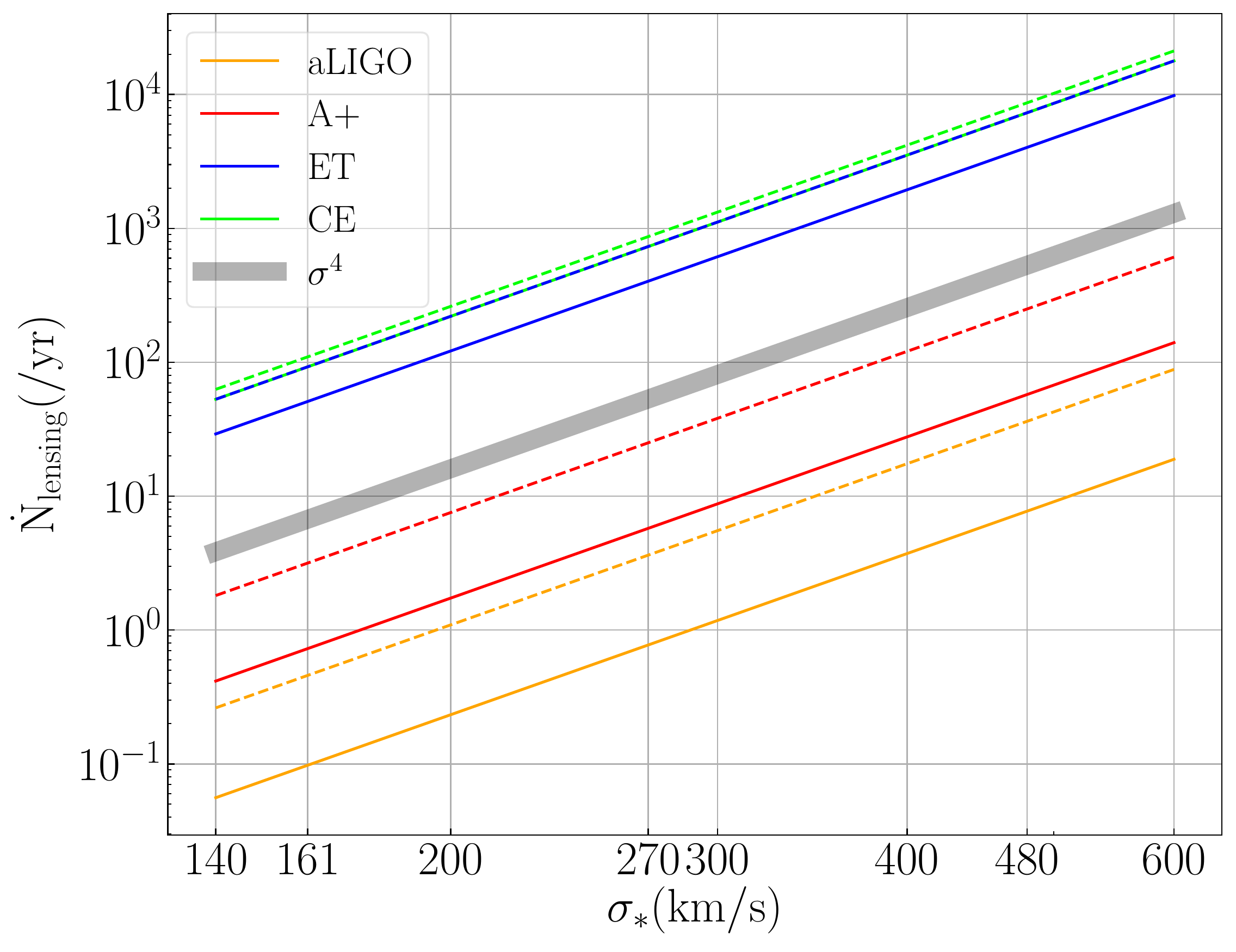}
\caption{{Predictions for the observed rate} of the primary images $\dot{N}_{\rm lensing, 1st}$ (dashed lines) and events with multiple images $\dot{N}_{\rm lensing, 2nd}$ (solid lines). Different colors represent different detectors. We set the SFR model to MD14 (\citet{2014ARA&A..52..415M}) with a minimum merger delay time $\Delta t_{\rm min}$ = 50 Myr, the galaxy number density to $\phi_*=8\times 10^{-3} h^3 \Mpc^{-3}$ \citep{2007ApJ...658..884C}, and use local BBH merger rate constrained by LIGO O2 $\mathcal{R}_0 = {64.9}_{-33.6}^{+75.5} \Gpc^{-3} \yr^{-1}$ \citep{2019arXiv190403187T}. $\dot{N}_{\rm lensing}$ is linearly dependent on $\phi_*$ and $\mathcal{R}_0$, and is proportional to $\sigma_*^4$. We also plot the grey line to mark the $\sigma_*^4$ trend.}
\label{Rlensing_sigma*}
\end{figure*}

As we can see from the calculation in Sections~\ref{subsec:tau}, \ref{subsec: BBHmerger}, and \ref{sec:lensing_rates}, when fixing the SFR, the local merger rate density $\mathcal{R}_{0}$, and the lens galaxy density $\phi_*$, the lensing event rate will be primarily determined by $\sigma_*$. 
We show the dependence of $\dot{N}_\text{lensing, 1st}$ and $\dot{N}_\text{lensing, 2nd}$ on $\sigma_*$ for both 2G and 3G detectors in Figure \ref{Rlensing_sigma*} and summarize the lensing event rate for the case of $\sigma_* = 161 \rm \mbox{km/s}$ in Table~\ref{tab:Rlensing_summary1}.

\begin{table*}
 	\centering
 	\caption{{Lensing event rate ($\dot{N}_\text{lensing}$, $\rm yr^{-1}$) assuming $\sigma_* = 161$ \mbox{km/s}  \citep{2007ApJ...658..884C}. We use SIE lens model for the calculation. We use MD14 as SFR model and set $\Delta t_{\rm min} = 50$ Myr. The 3 columns correspond to the lensing event rate of the primary image detected ($\dot{N}_{\rm lensing, 1st}$), the lensing event rate with at least 2 images detected($\dot{N}_{\rm lensing, 2nd}$), and the expected observed BBH merger event per year $\dot{N}_{\rm BBH}$ without considering magnification.}}
	
	\begin{tabular}{lccccr} 
		\hline
		 &  Primary image ($\dot{N}_{\rm lensing, 1st}$) & Multiply-imaged events ($\dot{N}_{\rm lensing, 2nd}$) & Unlensed BBH merger events ($\dot{N}_{\rm BBH}$)\\
 		\hline
 		aLIGO & 0.45 & 0.1  & $6.3\times 10^2$\\ 
 		A+ & 3.4& 0.7  & $3.7 \times 10^3$ \\ 
 		ET &93  & 51 & $1.2 \times 10^5$\\ 
 		CE & 110 & 92 & $1.5\times 10^5$ \\ 
		\hline
	\end{tabular}\label{tab:Rlensing_summary1}
\end{table*}

%

Figure \ref{Rlensing_sigma*} shows that both $\dot{N}_\text{lensing, 1st}$ and $\dot{N}_\text{lensing, 2nd}$ increase with $\sigma_{*}$, and are almost linearly dependent in log-space, $\log \dot{N}_{\rm lensing} \propto 4 \log \sigma_*$. This is mainly because $\tau(z) \propto \sigma_*^4$. By increasing the value of $\sigma_{*}$, the $\sigma$ of the whole lens galaxy population increases. Therefore, the overall cross-sections of lensing increases which results in higher $\dot{N}_\text{lensing}$. Moreover, $\dot{N}_\text{lensing}$ grows with the detector sensitivity since more sensitive detectors can observe a larger cosmological volume. We discuss the redshift distribution of lensed and unlensed events in more detail in Section \ref{sec: detectable cosmo volume}.

Table \ref{tab:Rlensing_summary1} summarizes the expected $\dot{N}_\text{lensing}$ using velocity dispersion $\sigma_* = 161$ \mbox{km/s} constrained from SDSS in the EM band \citep{2007ApJ...658..884C}. We assume the MD14 SFR model and a minimum delay time of $\Delta t_{\rm min}$ = 50 Myr. 
Our calculation shows that the contribution of the lensing events to the overall merger events is small, with a fraction of {$\lesssim 0.1 \%$}. The majority of the events that we detect should be unlensed events. Since we want to measure the time delay distribution, we are interested in events that have at least 2 images detected. The second column shows the estimation of $\dot{N}_\text{lensing, 2nd}$. According to these results, a statistical study of GW lensed events will happen with 3G detectors. Moreover, Appendix~\ref{app:Rz} shows results for two alternate SFR scenarios: MD14 SFR model with a different delay time of $\Delta t_{\rm min}$= 1 Gyr, and a different SFR density which is constant throughout the redshift evolution $\dot{\rho_{*}} = 0.004 M_{\odot}\Mpc^{-3} yr^{-1}$ with the same $\Delta t_{\rm min}$= 50 Myr.
{As an additional check, in Appendix~\ref{app:Rate lensing} we} compare the results in Table~\ref{tab:Rlensing_summary1} with the lensing event rates directly obtained from a lensing simulation. We find consistent results.

\subsection{Detectable cosmological volume increased by lensing magnification}
\label{sec: detectable cosmo volume}

\begin{figure*}
    \includegraphics[height=7cm,width=8.8cm]{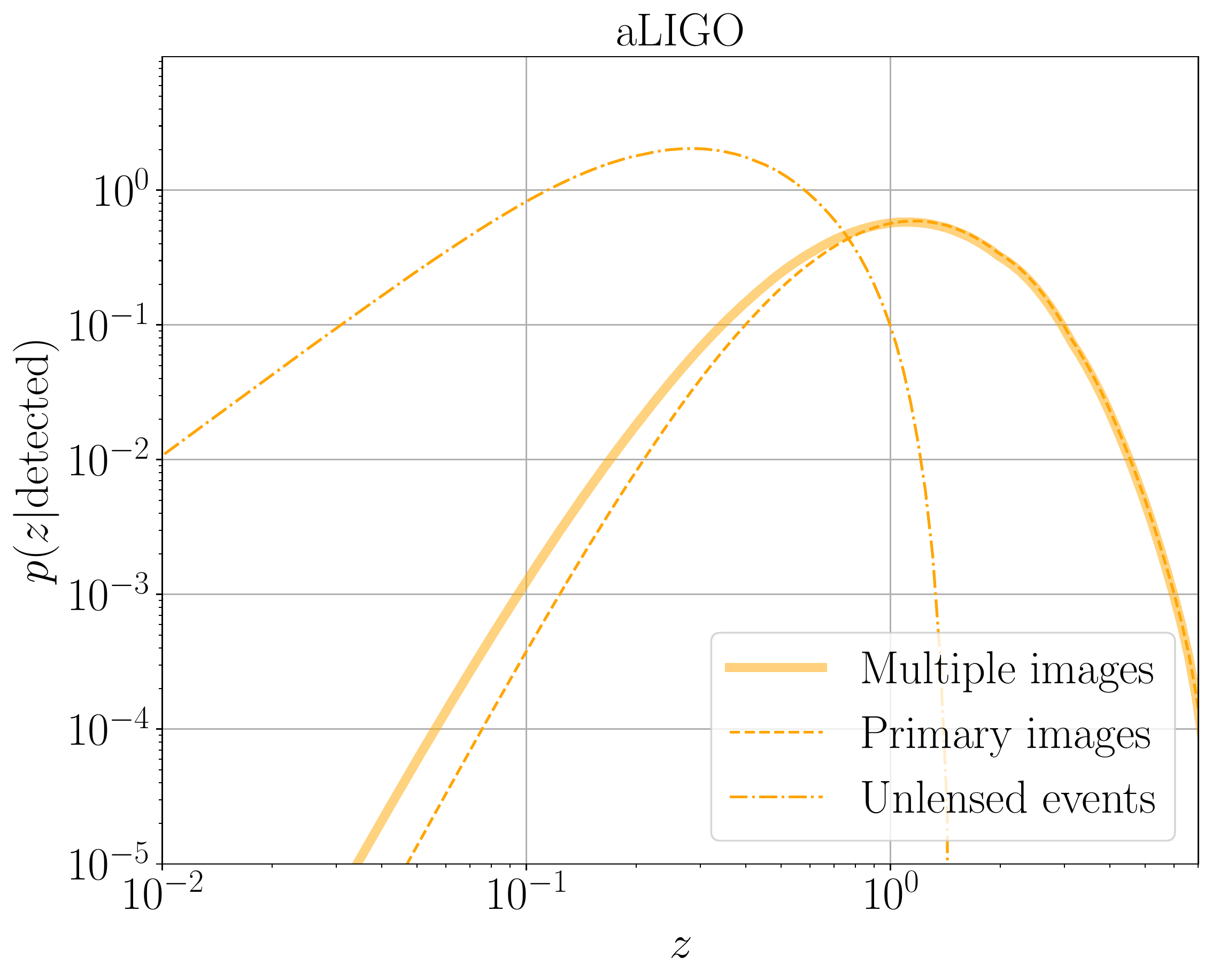}
    \includegraphics[height=7cm,width=8.8cm]{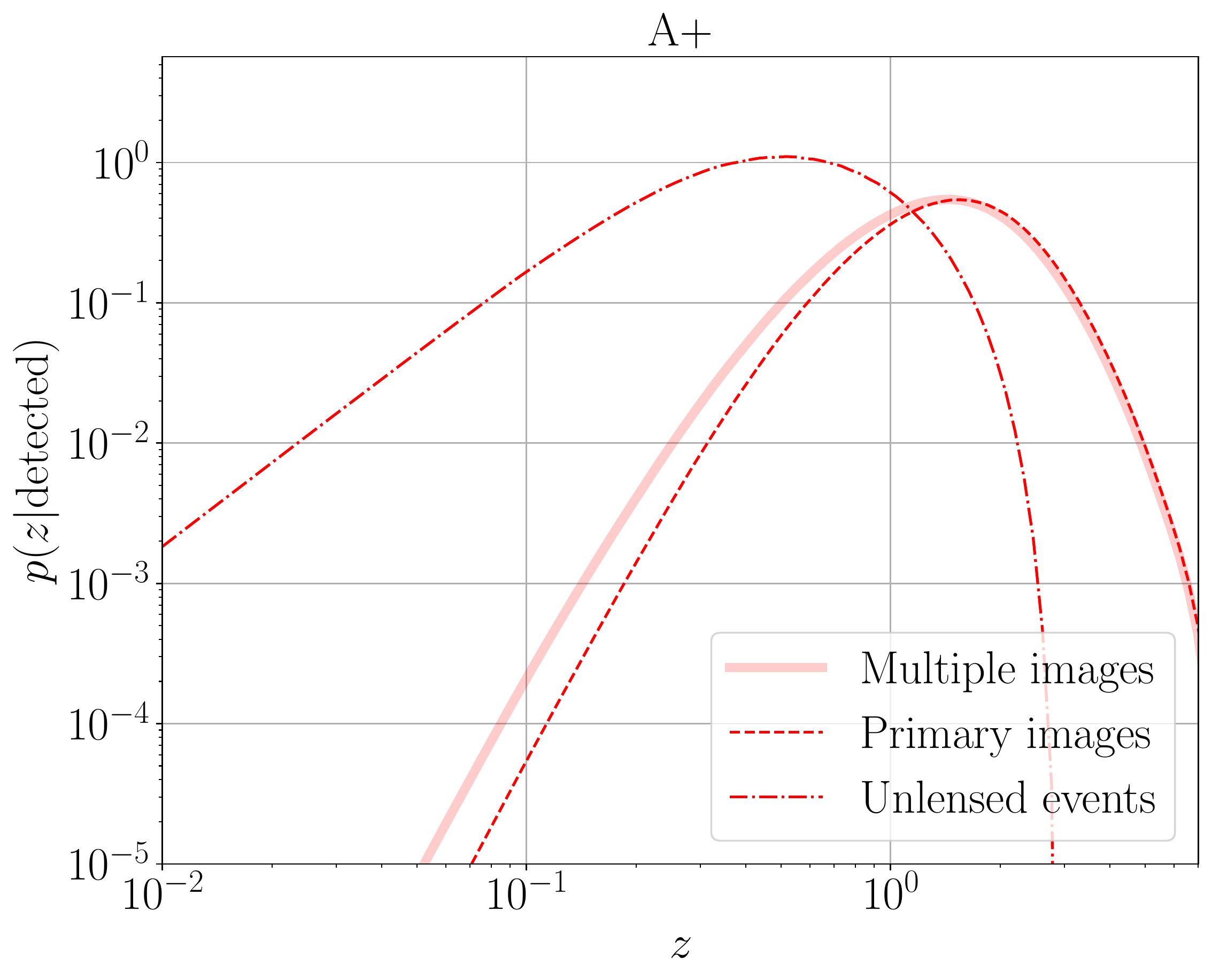}
    \includegraphics[height=7cm,width=8.8cm]{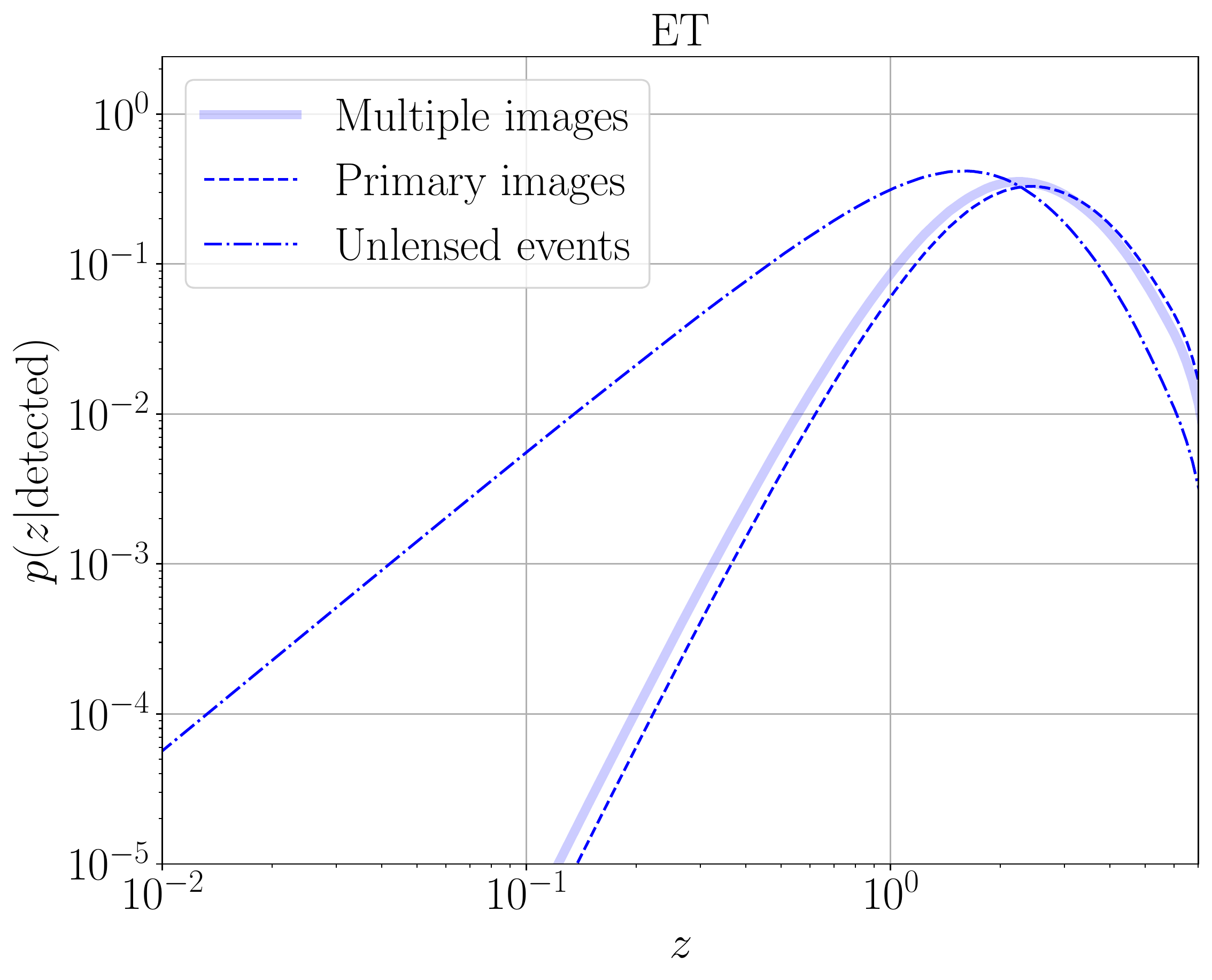}
    \includegraphics[height=7cm,width=8.8cm]{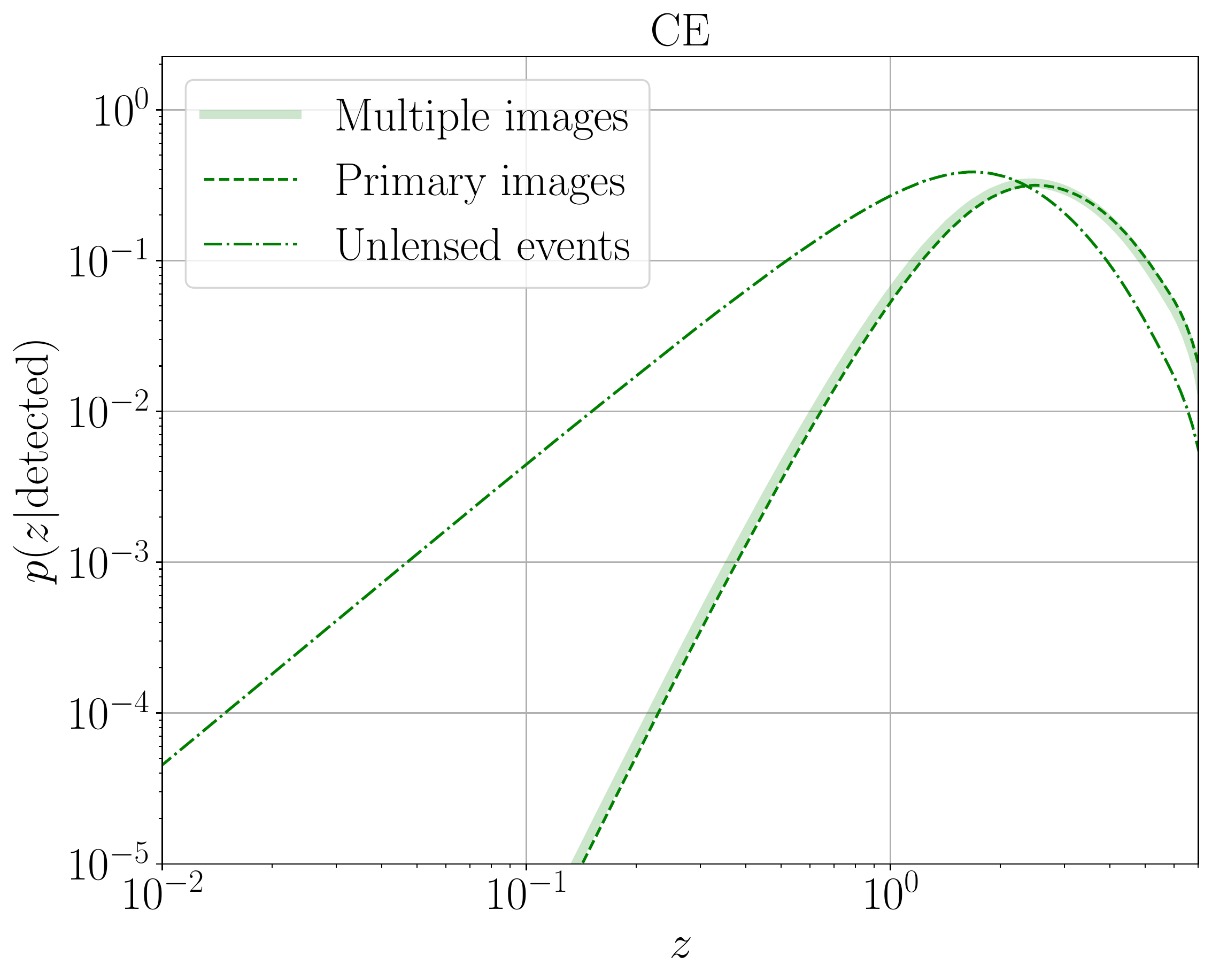}

	\caption{The redshift distribution of the unlensed images (dashed-dotted line, distribution of $\dot{N}_{\rm BBH}$), primary images (dashed line, distribution of $\dot{N}_{\rm lensing, 1st}$), and multiple lensed images (solid line, distribution of $\dot{N}_{\rm lensing, 2nd}$). We set $\sigma_{*} = 161 \rm \mbox{km/s}$ but the redshift distribution is not sensitive to $\sigma_*$. The characteristic redshifts within which 90 $\%$ of the events are included for different scenarios are summarized in Table \ref{tab:90perz}.}
    \label{fig:increase cosmo}
\end{figure*}

Since some of the BBH merger events at high redshift, which were previously too faint to be detected, can be magnified above the detection threshold, the detection volume of the GW detectors will increase due to lensing. Figure~\ref{fig:increase cosmo} shows the redshift distribution of the unlensed events, the primary images ($\dot{N}_{\rm lensing, 1st}$) which trace events with only one image detected, and the secondary images ($\dot{N}_{\rm lensing, 2nd}$) which trace the events with multiple images. Table \ref{tab:90perz} summarizes the characteristic redshift within which 90$\%$ of the events are included. The characteristic redshifts for aLIGO and A+ increase from $\sim 1$ to $\sim 3$, indicating that the detectable cosmological volume is drastically increased by lensing events, although as shown before, lensed events will only represent a very small fraction of the catalog. 
The same effect, however, is less significant for 3G detectors. The characteristic redshift for ET and CE increases from $\sim 4$ to $\sim 5$. This is mainly because: 1) GW sources at high redshift need higher magnification to be brought within the horizon than their low-redshift counterparts; and 2) since star formation generally peaks at $z\sim 2$ and drops at higher redshift, there are fewer BBH sources at $z>5$ as seen in Figure~\ref{fig:ddN/dzdt}, and therefore the increase in detectable volume due to lensing is slight. Moreover, the redshift distributions are insensitive to the change in $\sigma_*$, since we assume the lens galaxy population does not evolve with redshift in this paper. Therefore $\sigma_*$ mainly affects the {normalization of the lensing rate rather than the shape of the lensing optical depth with redshift}. We leave the incorporation of the redshift evolution of the lens galaxy population to future work.

\begin{table*}
 	\centering
 	\caption{Redshift within which 90 $\%$ of the unlensed events, $\dot{N}_{\rm lensing, 1st}$ and $\dot{N}_{\rm lensing, 2nd}$ are included.}
 	
	\begin{tabular}{lccccr} 
		\hline
		  & Unlensed events & Primary images & Multiple images \\
 		\hline
 		aLIGO & 0.9 & 2.8 & 2.7\\ 
 		A+ & 1.4 & 3.1 &  3.0 \\ 
 		ET & 3.8 & 5.1 &  4.8\\ 
 		CE & 4.1 & 5.2 & 5.1\\ 
		\hline
	\end{tabular}
	\label{tab:90perz}
\end{table*}

\subsection{Constraining galaxy populations using time delay distributions}
\label{sec: res delta t}

As mentioned in Section {\ref{subsec: sim}}, the lensing time delay $\delta t$ is directly related to the galaxy velocity dispersion $\sigma$ according to Equation \ref{eq:delt}, and thus can be used as a probe of the lensing population. 
The $\delta t$ distribution has a stronger dependence on $\sigma_*$ than the lensing rate, $\dot{N}_{\rm lensing}$, because $\dot{N}_{\rm lensing}$ is also affected by the number density of the lens galaxy ($\rm \phi_{*}$) and the BBH source population. Since different galaxy populations give different $\delta t$ distribution, we can use this fact to inversely constrain $\sigma_*$ from measuring the $\delta t$ distribution.
As described in Section \ref{subsec: sim}, we simulate the lensing process assuming a range of values for $\sigma_*$, and record the output time delay for each lensing system. Our simulations show that for aLIGO and A+ only a few lensing events will be detectable per year. Therefore, we only consider the measurements of the time delay distribution for 3G detectors, using ET as an example.

\begin{figure*}
\centering
\includegraphics[height=6cm,width=18cm]{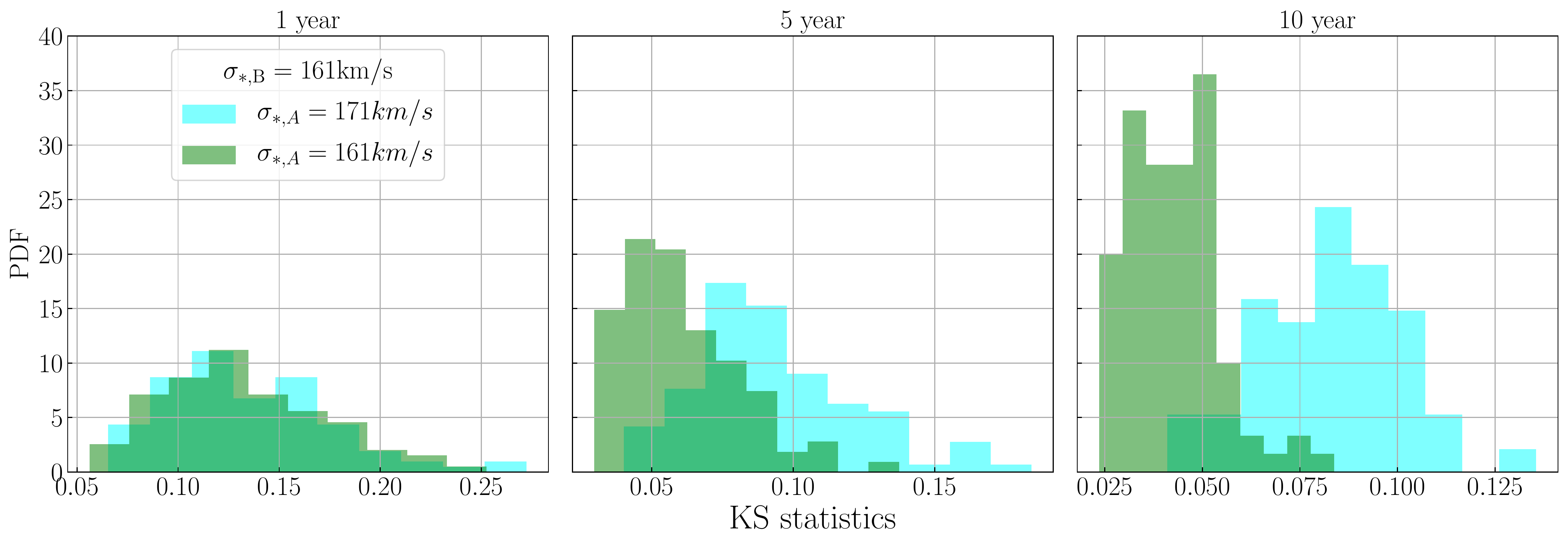}
\includegraphics[height=6cm,width=18cm]{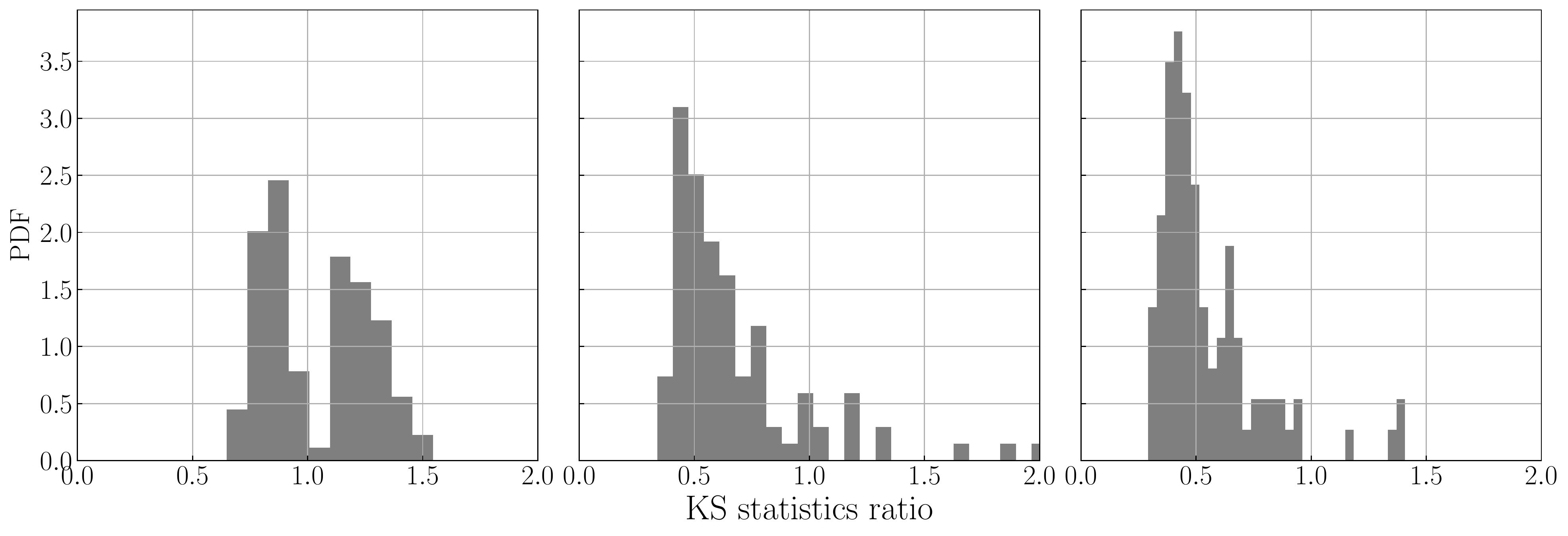}
\caption{KS statistics for 1 year, 5 years, and 10 years. {We use ET as the detector in this plot.} Upper panels: The PDFs of KS values from comparing samples generated from a model with $\rm \sigma_{*, B} = 161 \rm \mbox{km/s}$ and a model with $\sigma_{*, A} = 161 \rm \mbox{km/s}$ (i.e., $\rm KS(\sigma_{*, A} = 161 \rm \mbox{km/s}, \sigma_{*, B} = 161 \rm \mbox{km/s})$, green histogram) and from comparing samples generated from a model with $\rm \sigma_{*, A} = 171 \rm \mbox{km/s}$ and a model with $\sigma_{*, B} = 161 \rm \mbox{km/s}$ (i.e., $\rm KS(\sigma_{*, A} = 171 \rm \mbox{km/s}, \sigma_{*, B} = 161 \rm \mbox{km/s})$, cyan histogram). The 2 distributions diverge away from each other as the detection duration time increases, meaning we can better distinguish between models with $\sigma_{*} = 171 \rm \mbox{km/s}$ and with $\sigma_{*} = 161 \rm \mbox{km/s}$. Lower panels: The ratio of $\rm KS(\sigma_{*, A} = 161 \rm \mbox{km/s}, \sigma_{*, B} = 161 \rm \mbox{km/s})$ over $\rm KS(\sigma_{*, A} = 171 \rm \mbox{km/s}, \sigma_{*, B} = 161 \rm \mbox{km/s})$. The PDF area where the ratio is smaller than 1 is the probability of having correct inference which increases as we observe for a longer time.}
\label{fig:KS1}
\end{figure*}

Figure \ref{fig:KS1} provides an example of constraining galaxy properties using the time delay distribution. We generate 500 mock observation samples with $\sigma_{*, B} = 161$ \mbox{km/s} for 3 different observational durations (1 year, 5 years, and 10 years), and compare them with two theoretical models ($\sigma_{*, A} = 161$\,km/s and $\sigma_{*, A} = 171$\,km/s). The corresponding KS test statistics are denoted as KS($\sigma_{*, A} = 161$ \mbox{km/s}, $\sigma_{*, B} = 161$\,\mbox{km/s}) and KS($\sigma_{*, A} = 171$ \mbox{km/s}, $\sigma_{*, B} = 161$\,\mbox{km/s}) whose distributions are shown in the green and cyan histograms in the upper panels of Figure \ref{fig:KS1}. Since the green histogram shows the case where $\sigma_{*, A} = \sigma_{*, B}$, we use it as a reference. We can see that when ET only observes for 1 year, the green and cyan histograms almost overlap and hence it is hard to distinguish $\sigma_{*, A} = 161$\,\mbox{km/s} from 171\,\mbox{km/s}. However, as we gradually increase the observation time to 10 years, the green and the cyan histograms separate, indicating that a KS test is able to distinguish 161 \mbox{km/s} and 171 \mbox{km/s} using the observed time delay distributions.

The lower panel of Figure \ref{fig:KS1} shows the ratio of KS($\sigma_{*, A} = 161$ \mbox{km/s}, $\sigma_{*, B} = 161$ \mbox{km/s}) over KS($\sigma_{*, A} = 171$ \mbox{km/s}, $\sigma_{*, B} = 161$ \mbox{km/s}). As mentioned before, this ratio should always be smaller than 1 because the $\delta t$ distribution with $\sigma_{*, B} = 161$ \mbox{km/s} should be closer to theoretical model with $\sigma_{*, A} = 161$ than $\sigma_{*, A} = 171$ \mbox{km/s}. Nevertheless, due to the randomness of the sampling, and the limitation of the observation duration, the ratio may be bigger than 1 in some cases. This expectation is consistent with the lower panel of Figure \ref{fig:KS1} where we can see that the majority of the area of the histogram is smaller than 1. We can consider the area smaller than 1 as the probability of having correct inference for the true underlying $\sigma_*$. As expected, Figure \ref{fig:KS1} shows that when we increase the observation time, the area of the histogram at values < 1 gets larger while the area > 1 gets smaller, indicating that increasing observing time enhances the probability of having correct inference for the underlying lens model. 

\begin{figure*}
\centering
\includegraphics[height=10cm,width=15cm]{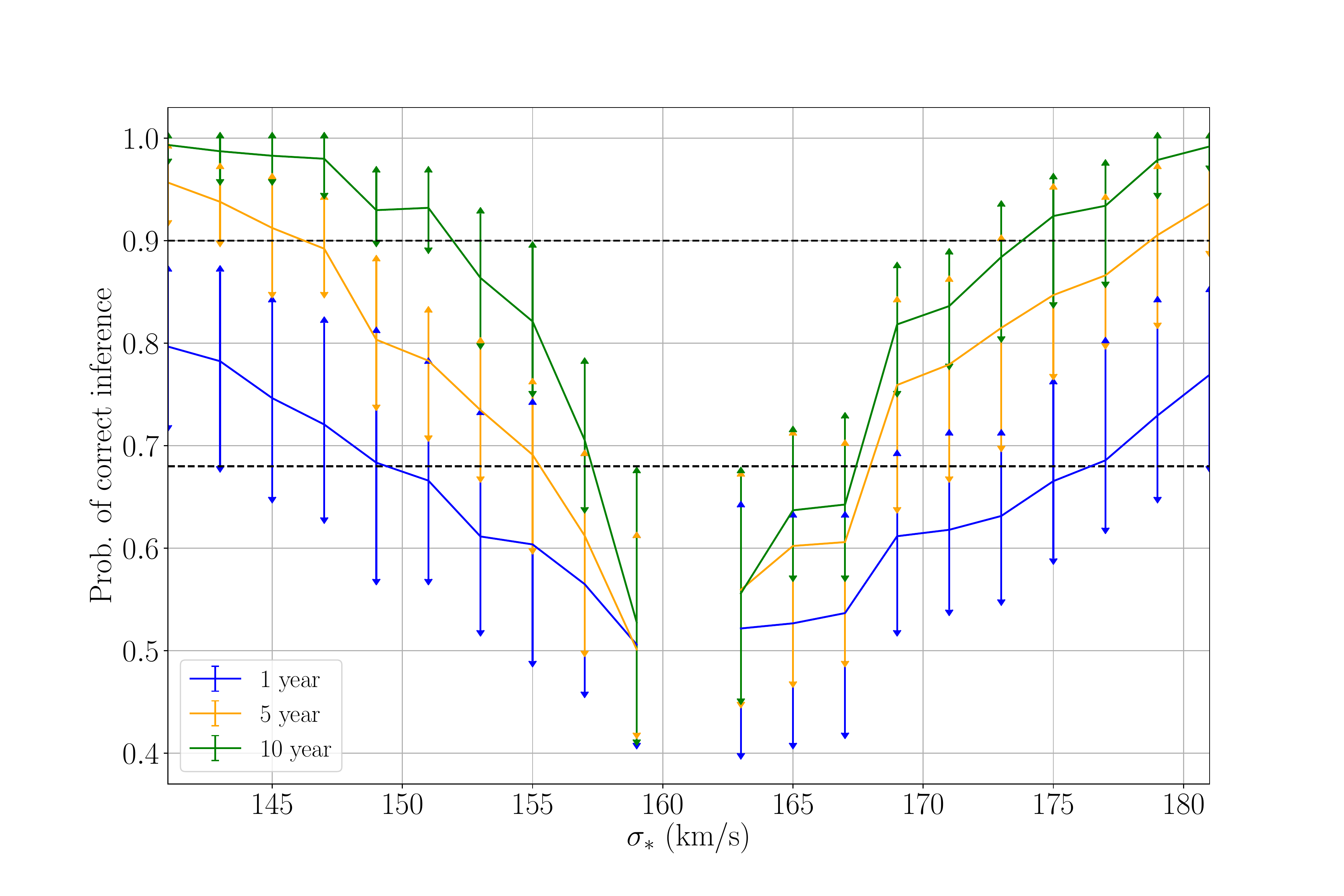}
\caption{Probability of correct inference as a function of $\sigma_{*, A}$ for the case of ET. We set the characteristic lens galaxy velocity dispersion of the mock observation sample to $\sigma_{*, B} = 161$ \mbox{km/s}. Solid line with different colors represent the average probability of correct inference for observation time from {1} year, {5} years, and {10} years. The bars show the minimum and the maximum probability of correct inference at the given $\sigma_{*, A}$. The black dashed lines mark the place where the probability of having correct inference equals to 68 $\%$ and 90$\%$. {We assume ET is always online within the observation duration time.}}
\label{fig:KS2_2}
\end{figure*}

We demonstrate how this procedure can be applied to other values of $\sigma_*$ in Figure \ref{fig:KS2_2} which shows $\sigma_{*, A}$ versus the probability of having correct inference assuming three different observation times: 1 year, 5 years and 10 years. Similar to Figure \ref{fig:KS1}, we use mock observational samples with $\sigma_{*, B} = 161$ \mbox{km/s} but now compare them with theoretical models in the range of $\sigma_{*, A} = $ 161 $\pm 20$ \mbox{km/s} with an increment of 2 \mbox{km/s}, instead of just $\sigma_{*, A} = 171$ \mbox{km/s}. We repeat this procedure 30 times and compute the average KS statistics and the maximum and minimum values as the bounds of the error bars. As expected, the probability of a correct inference improves when $\sigma_{*, A}$ is further away from the $\sigma_{*, B}$. {It is easier to distinguish models which are further apart.} {As shown in Figure \ref{fig:KS2_2}, for the case of $\sigma_{*, B} = $ 161 \mbox{km/s}, we can exclude $\sigma_* < $ 161-12 \mbox{km/s} and $\sigma_* > $ 161+16 \mbox{km/s} after 1 year of observation at 68 $\%$ confidence. Similarly, we are able to exclude $\sigma_* < $161-16 \mbox{km/s} and $\sigma_* > $ 161+18 \mbox{km/s} after 5 years of observation, and exclude $\sigma_* < $ 161-10 \mbox{km/s} and $\sigma_* > $ 161+14 \mbox{km/s} after 10 years of observation at 90 $\%$ confidence.}

\subsection{The effect of source population on the lensing rate}
\label{sec: res souce_pop_Rlens}

As described in Section \ref{sec:lensing_rates}, $\dot{N}_\text{lensing}$  depends not only on the lens population, but also the source population. The distribution of sources is determined by the particular BBH formation channel. In this section, we focus on how the assumptions related to binary formation evolution affect $\dot{N}_\text{lensing}$.

\begin{figure*}
\centering
\includegraphics[height=7cm,width=8.8cm]{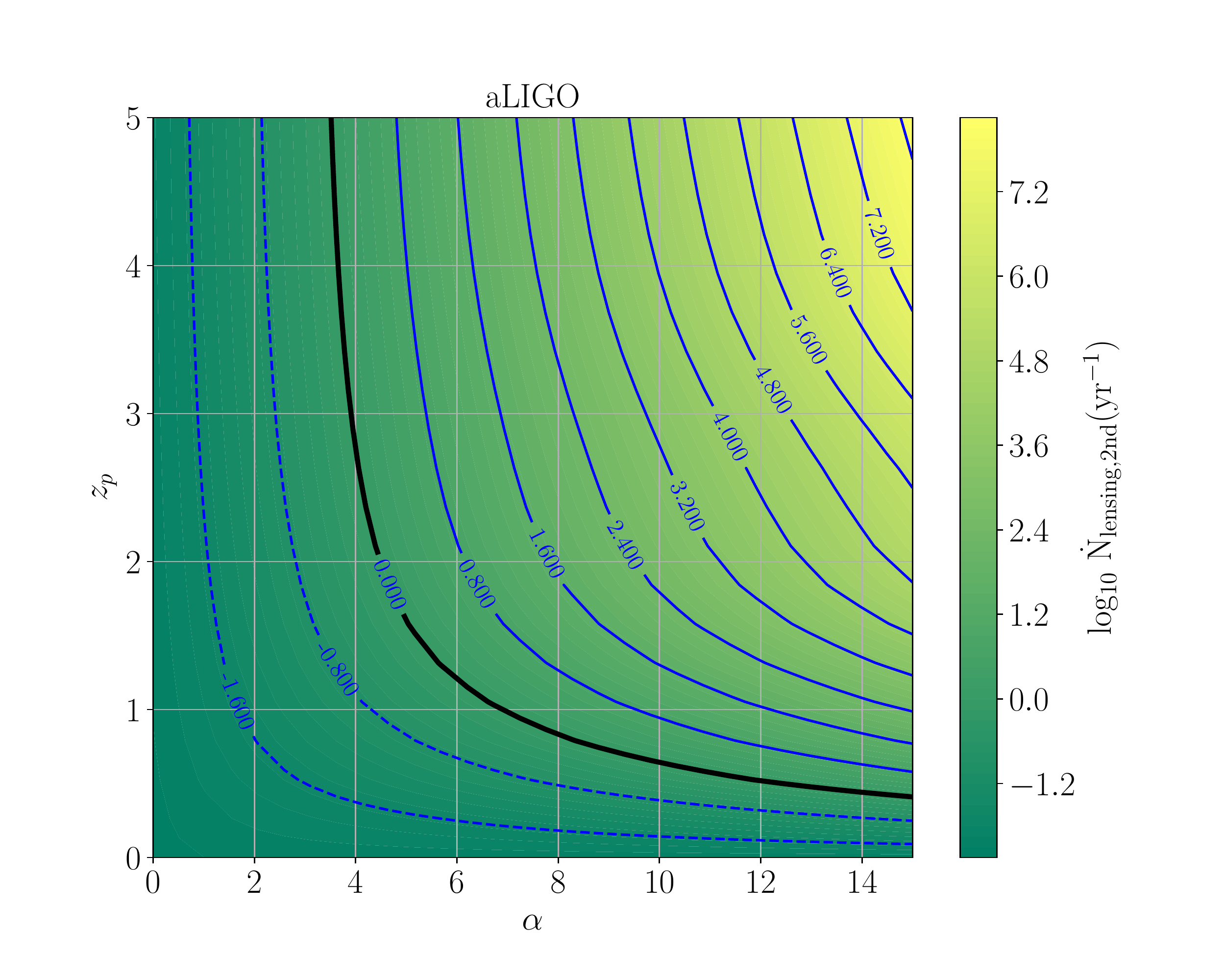}
\includegraphics[height=7cm,width=8.8cm]{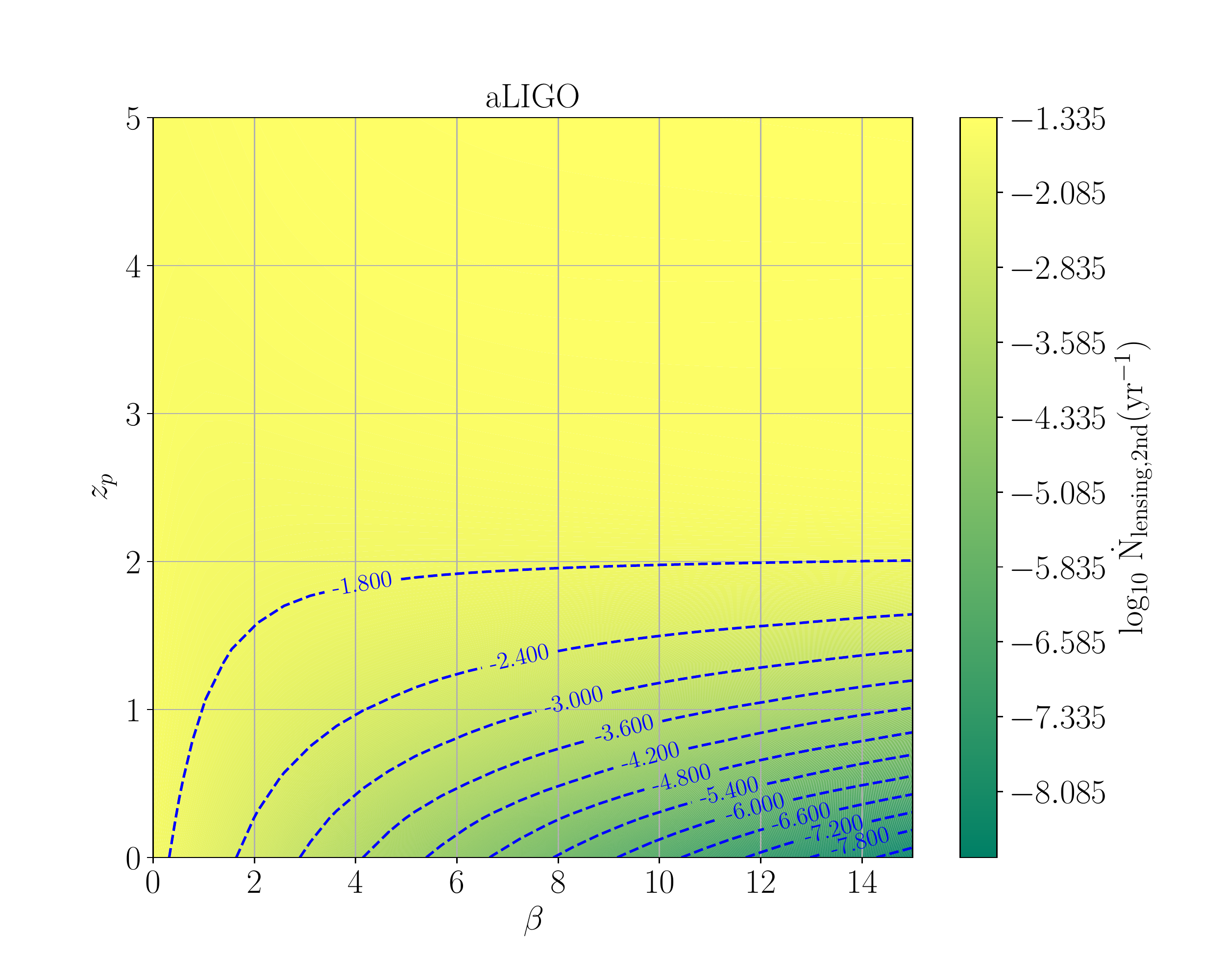}
\includegraphics[height=7cm,width=8.8cm]{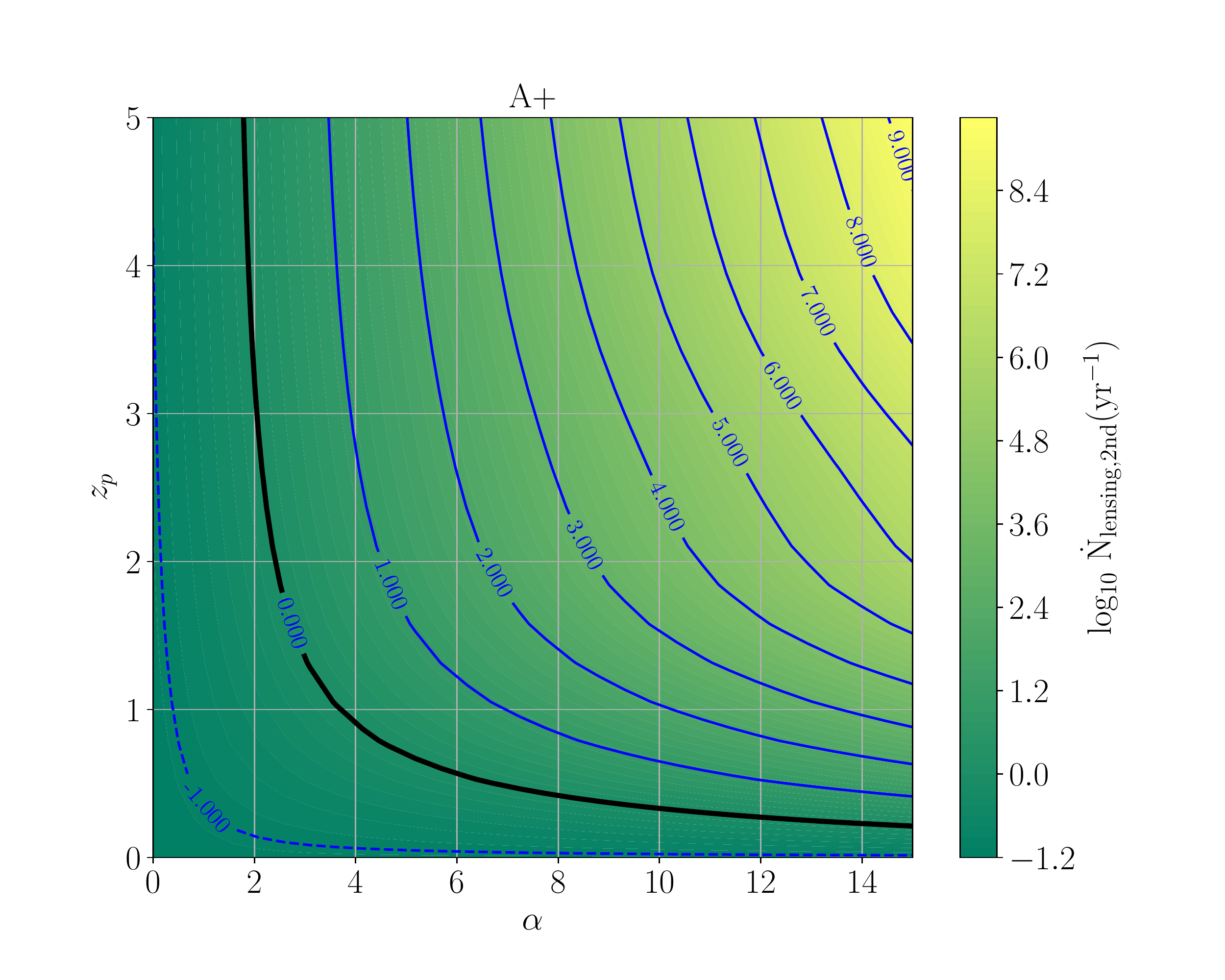}
\includegraphics[height=7cm,width=8.8cm]{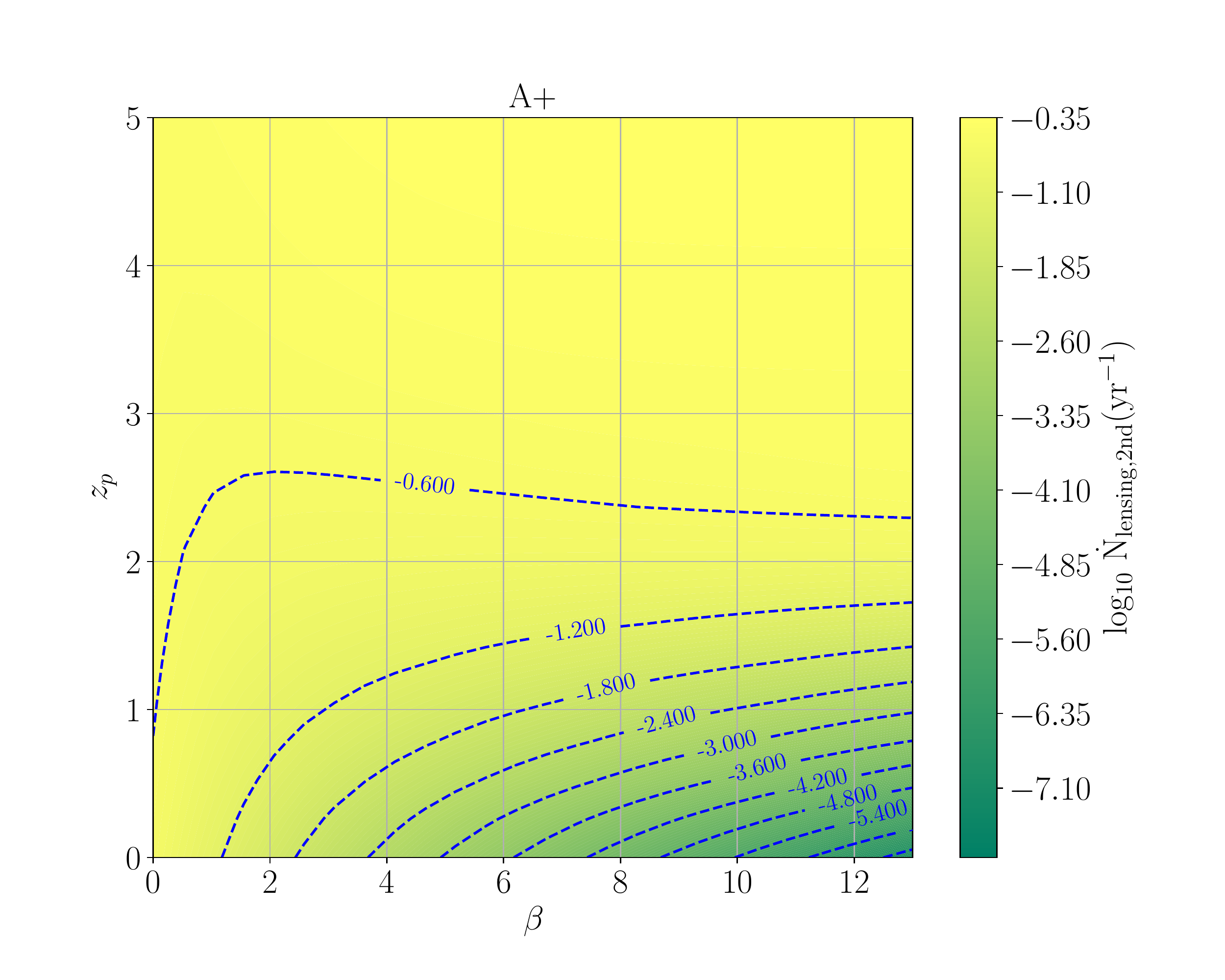}
\caption{Lensing event rate $\dot{N}_{\rm lensing, 2nd}$ distributions (number of the lensing pairs per year) for 2G detectors aLIGO and A+ still assuming $\sigma_{*} = 161 \rm \mbox{km/s}$. Contours are in $\rm log_{10}$ scale. Left column: $\dot{N}_{\rm lensing, 2nd}$ contour plot in $z_p-\alpha$ parameter space. We fix $\beta = 1$. The black solid line represents the parameter regime that will likely have 1 event per year. Right column: $\dot{N}_{\rm lensing}$ contour plot in $z_p-\beta$ parameter space. {We fix $\alpha = 1$ which is roughly at the peak of the constraint in O2 (See Figure 15 in \citet{2020arXiv201014533T}.)}}
\label{fig:contour2-2G}
\end{figure*}

\begin{figure*}
\centering
\includegraphics[height=7cm,width=8.8cm]{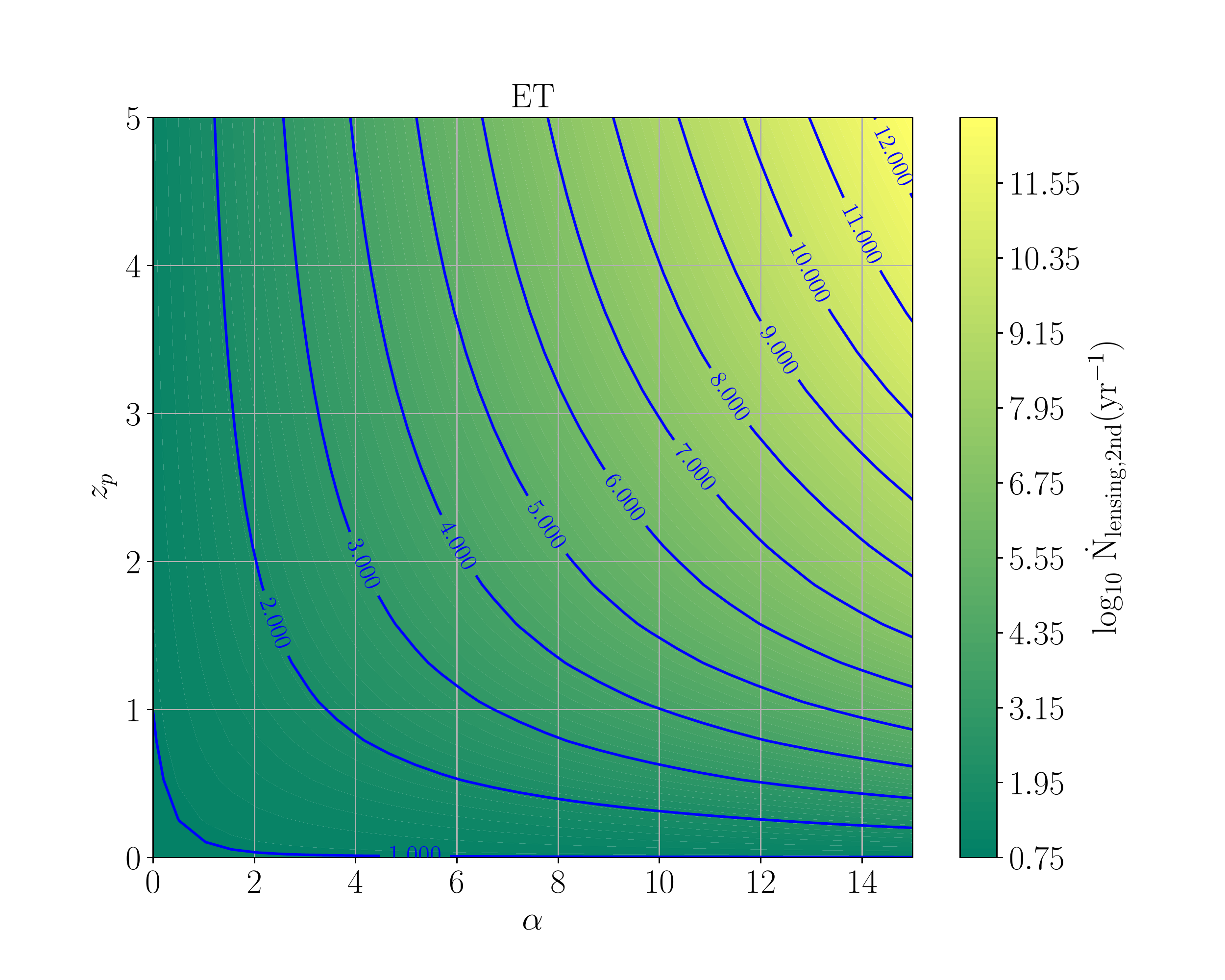}
\includegraphics[height=7cm,width=8.8cm]{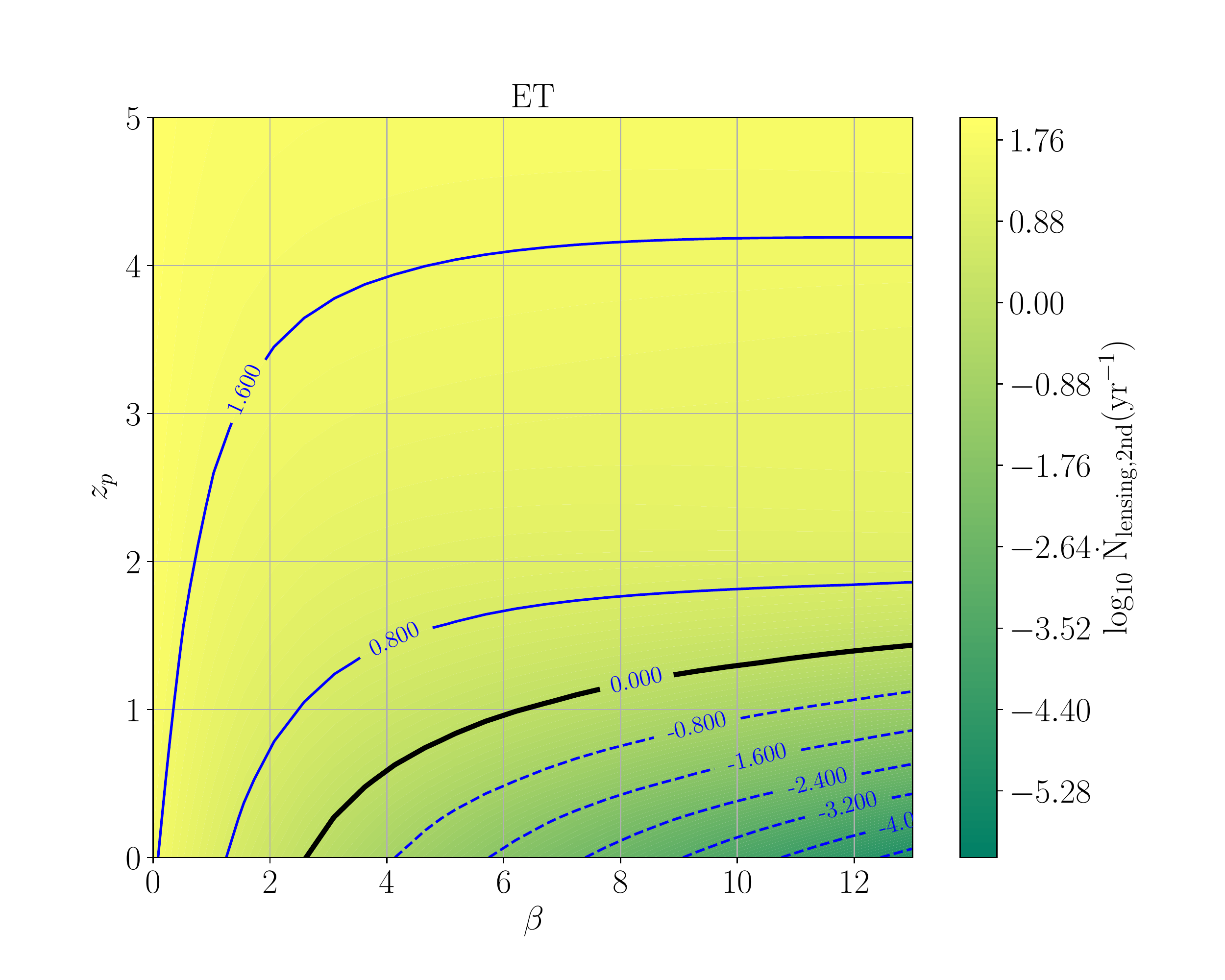}
\includegraphics[height=7cm,width=8.8cm]{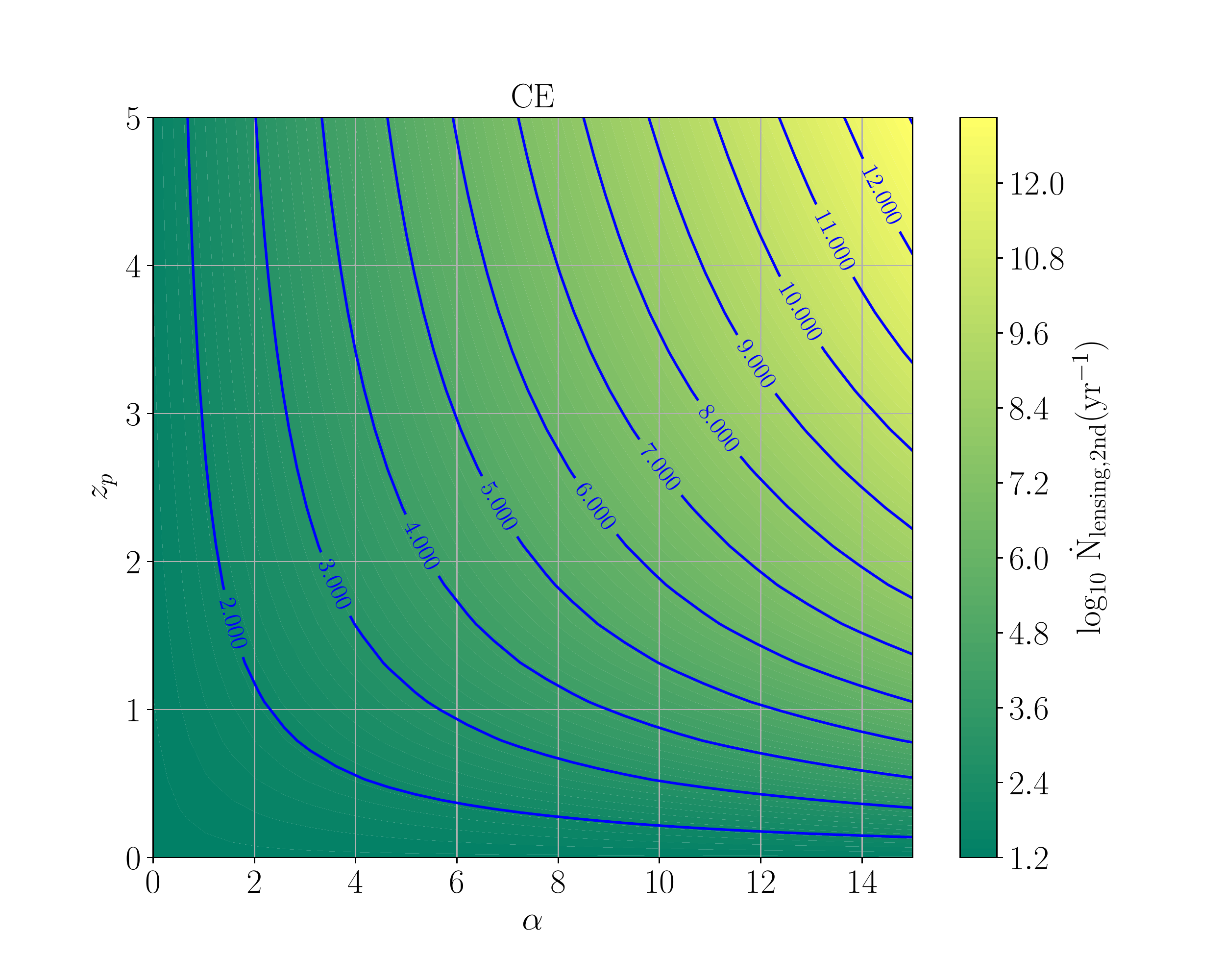}
\includegraphics[height=7cm,width=8.8cm]{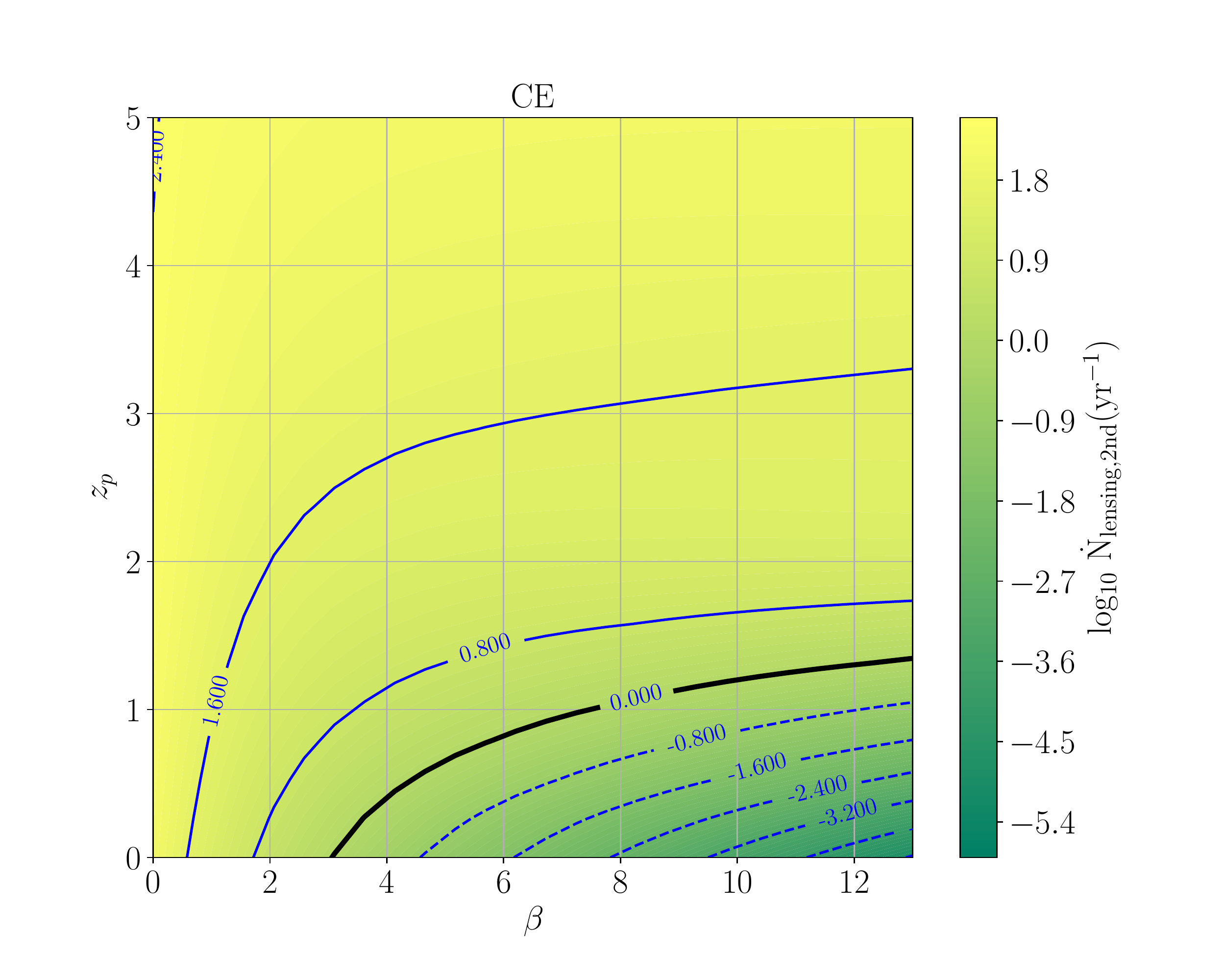}
\caption{Lensing event rate $\dot{N}_{\rm lensing, 2nd}$ (number of the lensing pairs per year) distributions for 3G detectors ET and CE assuming $\sigma_{*} = 161 \rm \mbox{km/s}$. Contours are in $\rm log_{10}$ scale. Left column: $\dot{N}_{\rm lensing, 2nd}$ contour plot in $z_p-\alpha$ parameter space. We fix $\beta = 1$. Right column: $\dot{N}_{\rm lensing, 2nd}$ contour plot in $z_p-\beta$ parameter space. {We fix $\alpha = 1$ which is roughly at the peak of the constraint in O2 (See Figure 15 in \citet{2020arXiv201014533T}.)}}
\label{fig:contour2-3G}
\end{figure*}

We calculate the rate of multiple images, $\dot{N}_{\rm lensing, 2nd}$, for different values of the BBH merger rate parametrization described in Equation \ref{eq:RBBH}, assuming a wide range of $\alpha$, $\beta$, and $z_p$. The results for 2G and 3G detectors are presented in Figure \ref{fig:contour2-2G} and \ref{fig:contour2-3G}, respectively. We can see the following features from the contour plot for aLIGO and A+: First, at constant $z_p$, $\dot{N}_{\rm lensing, 2nd}$ increases as $\alpha$ increases. This is because when increasing $\alpha$, the slope at low redshift gets steeper and the maximum value of $\dot{N}_{\rm BBH}$ becomes higher since the local merger rate is fixed. 
$\dot{N}_{\rm lensing, 2nd}$ increases as there are more BBH sources for larger $\alpha$. Second, at constant $\alpha$, $\dot{N}_{\rm lensing, 2nd}$ increases as $z_p$ increases. This is because the original BBH merger rate increases when we increase $z_p$ when fixing the local merger rate and the slope $\alpha$. Third, it is difficult for aLIGO to probe the region where $\dot{N}_{\rm lensing, 2nd} < \rm 1 yr^{-1}$ because it requires a long observation time to achieve a precise constraint on the lensing event rate. Finally, when $z_p > 5$, $\dot{N}_{\rm lensing, 2nd}$ stays roughly constant since aLIGO and A+ are mostly sensitive to BBHs at low redshift, thus any variations of $z_p$ at high redshift $z \gtrsim 6$ has minimal impact on the observed $\dot{N}_{\rm lensing, 2nd}$ for aLIGO.  For ET and CE, as shown in Figure \ref{fig:contour2-3G}, their detection ability is significantly deeper and the expected $\dot{N}_{\rm lensing, 2nd}$ is higher than for aLIGO and A+. Most of the region has $\dot{N}_{\rm lensing, 2nd} > \rm 1 yr^{-1}$, and thus 3G detectors are sensitive to a wider parameter space than 2G detectors.

Similarly, we can also draw contour plots for $\dot{N}_{\rm lensing, 2nd}$ as a function of $z_p$ and $\beta$, as shown in the right column of Figures~\ref{fig:contour2-2G} and \ref{fig:contour2-3G}.
We can see that at low $z_p$, $\dot{N}_{\rm lensing, 2nd}$ decreases with increasing $\beta$ and then stays constant. This is because when $\beta$ becomes large, the slope at higher redshift gets steeper but the general shape of the $\dot{N}_{\rm BBH}$ does not change significantly, hence $\dot{N}_{\rm lensing, 2nd}$ stays relatively constant. The contours for $\dot{N}_{\rm lensing, 2nd}$ become sparse when $z_p$ is high, due to similar reasons to those mentioned above regarding the sensitive cosmological volumes for different detectors. However, since we are considering BBHs formed following star formation, $z_p$ is unlikely to be greater than $z\sim 6$, so we only show results for cosmological distances within this.

These results show the potential for constraining the BBH source population with GW strong lensing observations. {Even the non-detection of strong lensing, and accompanying upper limits on the lensing event rate (e.g., $\dot{N}_{\rm lensing, 2nd} < 1 \rm yr^{-1}$) provide constraints on parameters {of the phenomenological model (Equation~\ref{eq:RBBH})} as shown in the contour plots in Figures~\ref{fig:contour2-2G} and \ref{fig:contour2-3G}.} 
We mark the contour where $\dot{N}_{\rm lensing, 2nd} = 1\,\mbox{yr}^{-1}$. {Alternatively, when fixing the formation scenario, strong lensing observations (or lack thereof) can be used to constrain the SFR and delay-time distribution, as we show in Appendix~\ref{app:Rz}.} As mentioned before, the lensing event rate in the region where $\dot{N}_{\rm lensing, 2nd} < 1 \rm yr^{-1}$ is difficult to constrain unless the observational duration is long. Therefore, 3G detectors provide better constraints on the source population than 2G detectors, not only because they are sensitive to higher redshift, but also because they have higher lensing event rates and the regions where $\dot{N}_{\rm lensing, 2nd} > 1\,\mbox{yr}^{-1}$ are wider. A realistic analysis of actual data would require a Bayesian population study so that all possible variables are varied at the same time. 
These analyses of individual events could be complemented with constraints from the stochastic background~\citep{Buscicchio:2020cij,Mukherjee:2020tvr}.

\section{Discussion $\&$ Conclusions}
\label{sec:discussion}

We have explored the use of strong lensing of gravitational-wave (GW) sources to study the distribution of galaxies and the population of binary black holes (BBHs) in the Universe. Unlike the case of strong lensing samples in the electromagnetic (EM) spectrum, GW astronomy offers the possibility of all-sky searches for lenses over a wide range of time delays, $\delta t$, with well characterized selection functions. Furthermore, GW sources do not suffer from dust extinction, and are completely unaffected by any sources of obscuration along the line-of-sight (except for gravitational effects, which are the whole point of this paper).
We argue that future samples of multiply-imaged GW events will provide a powerful probe with which to study properties of both the lenses and the sources.

We calculate the lensing event rate $\dot{N}_{\rm lensing}$ assuming different lens galaxy and source populations.  
We consider the number of {detected} lensing pairs per year, $\dot{N}_{\rm lensing, 2nd}$, where \emph{at least} 2 images of a lensing system are detected.
We adopt a magnification distribution for the second brightest image, $P(\mu)_{\rm 2nd}$, for the calculation of $\dot{N}_{\rm lensing, 2nd}$ and discuss alternate definitions of the magnification distribution in Section \ref{sec:lensing_rates}. We summarize the results in Table \ref{tab:Rlensing_summary1} and Figure \ref{Rlensing_sigma*}. {In general, our prediction for the lensing event rate is consistent with the current non-detection of lensing events \citep{2019ApJ...874L...2H, 2020PhRvD.102h4031M, 2020arXiv200712709D,Abbott:2021iab}. }

We find that for typical lens parameters, the expected number of lensing pairs per year is about {0.1} $\rm yr^{-1}$ for aLIGO and {$\sim$ 1} $\rm yr^{-1}$ for A+. We expect hundreds of events will be detected with 3G GW detectors such as ET and CE. The overall fraction of the lensing events relative to unlensed BBH merger events is $\lesssim 0.3\%$, thus the majority of the events are unlensed ones. Yet detecting even just a few strongly lensed GW systems will provide useful information.

We stress that our results are subject to our assumptions about the lens and source populations.  
We demonstrate the dependence of the lensing event rate on the galaxy population, characterized by $\sigma_*$, in Figure~\ref{Rlensing_sigma*}. We also test the dependence of the strong lensing rate on the source population in Figure \ref{fig:contour2-2G} and Figure \ref{fig:contour2-3G}, using a parameterization in terms of $\alpha$, $\beta$, and $z_{p}$ in Equation~\ref{eq:RBBH}. We mark the region with a black solid line where we expect the detection of 1 lensing pair {per year}. These plots demonstrate that both detection and non-detection of lensing events will provide valuable information on the lens and source population.

In addition, by performing lensing simulations, we show that the distribution of time delays between multiply-imaged events provides helpful information to constrain the population of lenses, and is especially sensitive to the characteristic galaxy velocity dispersion, $\sigma_*$ defined in the Schechter function in Equation \ref{eq:Collettpdf}. We show that 3G detectors such as ET could constrain $\sigma_*$ using the shape of the time delay $\delta t$ distribution to a precision of {17 $\%$} at $68\%$ probability after 1 year of observation, and {$\sim$15$\mbox{--}$21 $\%$} at $> 90\%$ probability {after $\sim 5\mbox{--}10$ years of observation.} 
The time duration to achieve this precision may differ from this prediction, since it depends on the actual lensing event rate, $\dot{N}_{\rm lensing}$, which depends on $\sigma_*$, the galaxy density $\phi_*$, and the source population $\mathcal{R}_0$. Our main focus in this paper is to show the potential of using the $\delta t$ distribution to constrain the lens population; the precise constraints will depend on the lensing rate.

We emphasize that this paper adopts a highly simplified model. Our aim is to provide a demonstration of the tremendous potential of GW lensing, rather than the definitive analysis of lensing constraints. 
Our analysis could be extended in several ways. First, we have assumed that the lens galaxy population does not evolve with redshift. However, we know that galaxy populations do evolve over cosmological time in the real Universe. For example, the Schechter function describing the velocity dispersion distribution in Equation \ref{eq:Collettpdf} will also be a function of redshift. The redshift evolution of the lens population is something that would be accessible with GW lensing. 
For example, \cite{2018MNRAS.480.3842O} takes into account the redshift evolution of the galaxy density $\phi_*(z)$ based on the result from hydrodynamic simulation. The number density of the galaxies with $\sigma >~150$ \mbox{km/s} at $z=6$ is about an order-of-magnitude lower than their low-redshift counterparts. 
Changing the density evolution of galaxies will change the rate of strong lensing $\dot{N}_{\rm lensing}$ as a function of redshift, {and this is another interesting quantity which lensing populations will be able to probe.} 
We note that this evolution will nonetheless preserve the shape of the magnification distribution (see Figure 11 in \citet{Wang:2021kzt}). Including these effects will yield more realistic estimation of the strong lensing event rate.
The inclusion of lensing analyses from detailed cosmological simulations may also offer significant improvements in estimates of lensing rates and statistics. 
Nonetheless, aLIGO and A+ are more sensitive to events in the local Universe and hence the calculations are not significantly affected for these cases. {Moreover, there is a large uncertainty in our understanding of the distribution of lenses at high redshift. GW astronomy may shed lights on these measurements and provide unique insights on the dark matter distribution at those redshifts.} 

This work could also be extended by considering different sources. Here, we have focused on stellar-mass BBHs from canonical astrophysical origin, characterized by the star formation rate and delay-time distributions. However, there are alternative models which may also be probed. For example, BBHs forming directly from population-III stars could lead to high redshift GW events, with a merger rate history peaking around $z\sim10$ \citep{Kinugawa:2014zha,Hartwig:2016nde,Belczynski:2016ieo}. These events are likely to be strongly lensed due to their high redshifts {and would be perfect targets for 3G detectors \citep{Ng:2020qpk}}. 
Similarly, primordial BBHs could be another source of GWs at high redshift. Such primordial BHs can be formed due to the gravitational collapse of primordial density fluctuations in the early Universe \citep{Sasaki:2018dmp}. They could form binaries soon after their formation and merge at high redshift, hence are also likely to be lensed. One could potentially use the observation/non-observation of strongly lensed GW events to constrain PBH models and star formation models of the early cosmic epoch in general. {Although in this work we have focused on the lensing rates and their redshift evolution, future GW detectors will also provide information about the mass distribution of the strongly lensed events. This would be a key factor to disentangle the origin of the lensed events and if they constitute the same population as the unlensed binaries.}

{Below the frequency band of ground-based detectors, more massive binaries could also be strongly lensed. Promising sources are super-massive black hole bianries} 
(SMBBH) whose population could extend to redshift larger than 10 \citep{2002MNRAS.331..805H, Sesana_2005, 2007MNRAS.377.1711S, 2009PhRvD..80f4027K}. 
There are, however, lots of uncertainties in the actual population of SMBBH which will consequently affect the strong lensing event rate. Typical scenarios expect tens of event per year, although only a few of them might end up being multiply imaged within a 5-year LISA mission \citep{2010PhRvL.105y1101S}. 
{In addition, if present in nature, intermediate mass binary black holes would constitute a perfect target for lensing studies since LISA will hear them across the history of the Universe. 
Our analysis could be naturally applied to these LISA populations.}

{Our simplified lens model could be enriched in a few ways.} {Larger mass distributions like galaxy clusters \citep{2018IAUS..338...98S} or more compact lenses like black holes \citep{Virbhadra:1999nm, 2019PhRvD.100b4018G,2020PhRvD.101d4031G} may also play a role in strong lensing.} {GW lensing is sensitive to an enormous range of scales, ranging from time delays of fractions of a second up to many years, spanning over 10 orders of magnitude. EM lensing surveys are often significantly more restricted due to selection effects of the sample.} GW lensing can also potentially discover low-surface brightness sub-halos \citep{2020AJ....159...49D, 2020ApJ...901...58O}. We leave a comprehensive analysis of the full distribution of lenses for future work. 
Likewise, a fully detailed calculation of the BBH rate at high redshift should take into account weak lensing, which could potentially modify the magnification distribution and bias standard siren measurements \citep{Holz:1997ic,Holz:2004xx,2010PhRvD..81l4046H}. Weak lensing of GWs is also discussed in \citep{2020PhRvD.101j3509M,2020MNRAS.494.1956M}.
Microlensing of gravitational waves is another important topic \citep{Cheung:2020okf} which could be incorporated in future work.

It is to be noted that the magnification distribution can be measured for standard candles such as Type Ia supernovae, and that this also provides a unique lensing constraint~\citep{1999A&A...351L..10S,2018PhRvL.121n1101Z}. In addition, the magnification maps of such sources can be used to probe cosmology~\citep{2006ApJ...637L..77C}. GW sources offer a direct measurement of distance, since their intrinsic luminosities are calibrated by general relativity to make them standard sirens~\citep{1986Natur.323..310S,2005ApJ...629...15H}. However, without an independent estimate of the redshift, lensing magnification is completely degenerate with distance. This degeneracy might be broken by finding an optical counterpart~\citep{2006PhRvD..74f3006D,2017Natur.551...85A}, or by using known properties of the mass distribution~\citep{2019ApJ...883L..42F,2021ApJ...909L..23E}.
In addition to being another powerful constraint of the lens population, a combined analysis of both the time delay distribution and the magnification distribution may provide simultaneous constraints on $\sigma_*$ and $q_g$. The magnification ratio, $\mu_2/\mu_1$, will be measureable from a population of lensed GW sources. This could provide a potentially interesting probe of the axis ratio, or the ellipticity distribution of lens galaxies, {as discussed in Appendix~\ref{app:mag ratio}}. Gamma-ray bursts and fast radio bursts also provide interesting source populations for lensing studies~\citep{1999ApJ...510...54H, 2016PhRvL.117i1301M, 2017ApJ...842...35C}, in a similar fashion to the GW sources discussed in this paper.

{In addition to astrophysics, statistical studies of the population of strongly-lensed gravitational waves will provide valuable information about fundamental physics. In particular, the optical depth and time delay distribution computed in this work assumes the validity of general relativity. Modified theories of gravity could, for example, produce additional echoes to each lensed images \citep{Ezquiaga:2020dao} and alter the net time delay distribution. {Alternate models of gravity might also change the growth of structure in the universe and hence the matter distribution of the lenses~\citep{2019ARA&A..57..335F}, which would leave discernible imprints on the lensing populations.} Thus, our methodology could be extended to directly test general relativity.}

As the sensitivity of GW detectors improves, the detection of strong lensing of GW sources is inevitable. Statistical studies of this lensed population will provide important probes of the properties of the lenses as well as novel constraints on the properties of GW sources.
These strong lensing events will open a new door to exploration of the inhomogeneous Universe.

\acknowledgments
We are thankful to Thomas Callister, Chihway Chang, Will Farr, Maya Fishbach, Mike Gladders, Otto Hannuksela, {Anupreeta More}, and Haowen Zhang for useful discussions. {We also thank Renske Wierda and Otto Hannuksela for pointing out that the sign of $B$ in Eq.~(\ref{eq:p_qg}) should be negative.}
{FX and DEH are supported by NSF grants PHY-2006645 and PHY-2011997.}
JME is supported by NASA through NASA Hubble Fellowship grant HST-HF2-51435.001-A awarded by the Space Telescope Science Institute, which is operated by the Association of Universities for Research in Astronomy, Inc., for NASA, under contract NAS5-26555. JME and DEH are also supported by the Kavli Institute for Cosmological Physics through an endowment from the Kavli Foundation and its founder Fred Kavli. {DEH~gratefully acknowledges the Marion and Stuart Rice Award.}
This material is based upon work supported by NSF LIGO Laboratory which is a major facility fully funded by the National Science Foundation.

\clearpage
\appendix
\section{Optical depth for different lens models}
\label{app:tau}

\begin{figure*}
\centering
    \includegraphics[height=8cm,width=10cm]{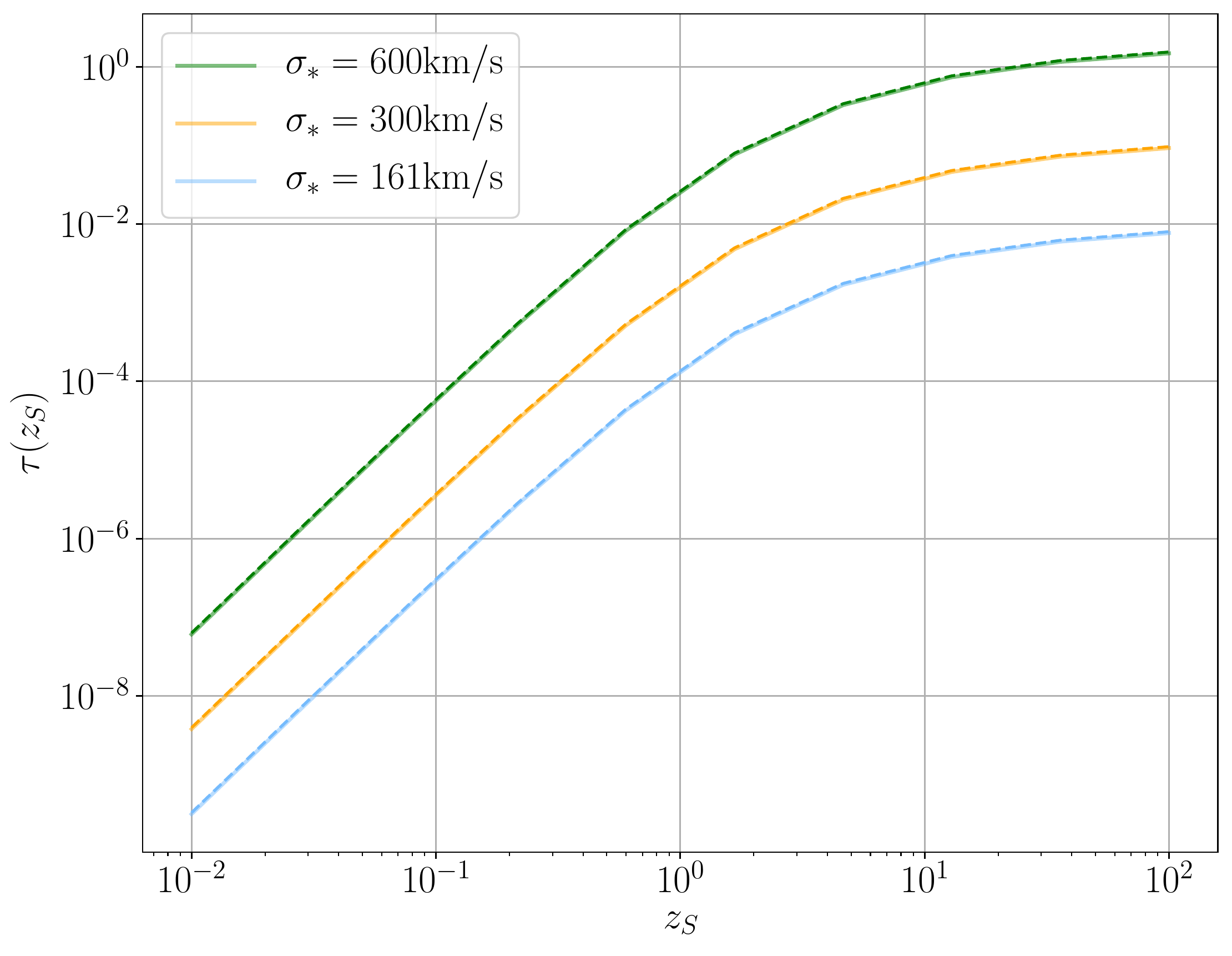}
	\caption{The optical depth as a function of source redshift $z_{\rm S}$. Solid and dashed lines represent $\tau$ assuming SIE model and SIS model respectively. Different colors represent different $\sigma_*$.}
    \label{fig:tau sis sie}
\end{figure*}

{In this appendix, we compare the optical depth $\tau$ derived using a Singular Isothermal Ellipsoid (SIE) model as in the main text ($\tau$) with the Singular Isothermal Sphere (SIS) model ($\tau_{\rm SIS}$).} 
The cross section of the SIS lens model is simply given by the Einstein radius $\theta_E$ of the lensing system in {Equation \ref{eq:theta E}}. Apart from the geometrical configuration of the source-lens system, the Einstein radius is fully determined by the galaxy velocity dispersion $\sigma$.

Similarly as in Section \ref{subsec:tau}, the differential optical depth is given by:
\begin{align}
 \frac{d\tau_{\rm SIS}}{dz_{\rm L}}  
 &\quad= {\int_{\sigma_{\rm min}}^{\sigma_{\rm max}} \frac{dV_c}{\delta\Omega dz_{\rm L}}n(z_{\rm L},\sigma)\pi \theta_E^2(z_s, z_L, \sigma)\ d\sigma} \\
 &\quad= {\int_{\sigma_{\rm min}}^{\sigma_{\rm max}} 16\pi^3 \frac{c(1+z_{\rm L})^2}{H(z_{\rm L})} \left(\frac{D_{\rm L}D_{\rm LS}}{D_{\rm S}}\right)^2\left(\frac{\sigma}{c}\right)^4 \phi(\sigma|z_L) d\sigma} \nonumber
\end{align}
where $\sigma_{\rm min}$ and $\sigma_{\rm max}$ are the lower and upper bound of $\sigma$. We substitute the differential comoving volume per solid angle in the second line. 
When setting $\sigma_{\rm min} = 0$ and $\sigma_{\rm max} = \infty$ and fixing the Schecter function for the number density of the lenses $\phi(\sigma|z_L)$, the expression of $\tau_{\rm SIS}$ can be integrated analytically \citep{2018arXiv180707062H}:
\begin{align}
\tau_{\rm SIS} = 16 \pi^3 (\frac{\sigma_*}{c})^4 \frac{\Gamma (\frac{4+\alpha}{\beta})}{\Gamma (\alpha/\beta)} \frac{n D_c(z_s)^3}{30}\,.
\label{eq:tau3}
\end{align}
We can generalize the above expression to arbitrary integration bounds $\sigma_{\rm min}$ and $\sigma_{\rm max}$. Again, this can be integrated analytically. We obtain: 
\begin{equation}
\begin{aligned}
\tau_{\rm SIS}(z_S) = 16 \pi^3 (\frac{{\sigma_*}^4}{c}) \frac{n {D_c(z_s)}^3 }{30} \frac{\Gamma(\frac{\alpha+4}{\beta}, \frac{\sigma_{\rm max}}{\sigma_*}) - \Gamma(\frac{\alpha+4}{\beta}, \frac{\sigma_{\rm min}}{\sigma_*})} {\Gamma (\frac{\alpha}{\beta})}\,.
\label{eq:tau general}
\end{aligned}
\end{equation}

When comparing this calculation with the one in the main text, we find that the ratio of optical depths is $\tau/\tau_{\rm SIS} \approx {0.96}$ and almost stays constant throughout the redshift range $z=1 \sim 100$. 
This can be seen in Figure \ref{fig:tau sis sie} where we plot both optical depths for different values of $\sigma_*$. The product $\tau_{\rm SIS} \times {0.96}$ gives a good approximation to the SIE optical depth $\tau$ with a difference of only $\sim 0.5\%$, when $\sigma_{\rm min} = 0$, $\sigma_{\rm max} = \infty$. The factor of 0.96, however, does not apply to other scenarios.

\begin{figure*}
\centering
    \includegraphics[height=8cm,width=10cm]{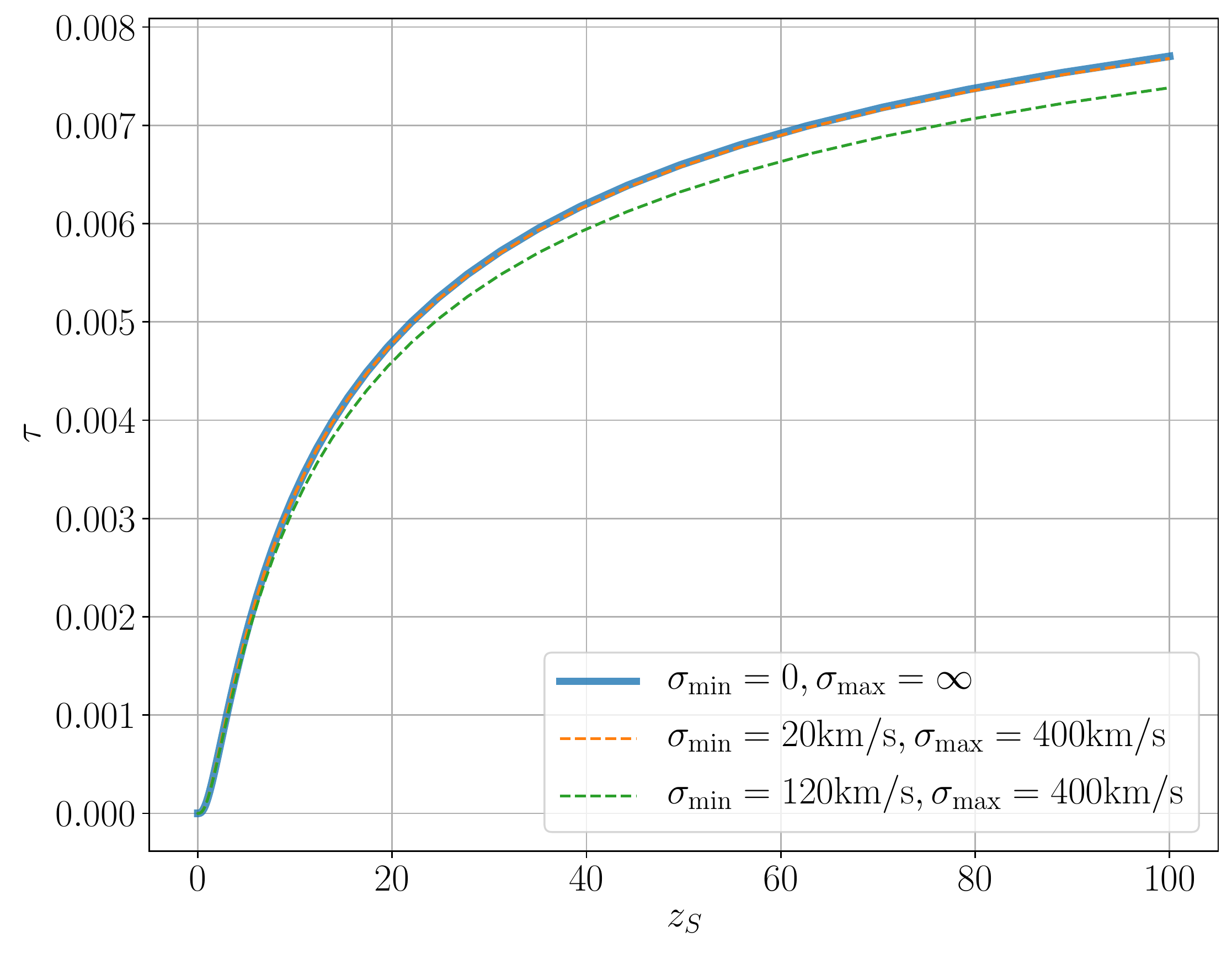}
	\caption{SIE optical depth assuming different upper and lower bounds for the velocity dispersion $\sigma$. The sky-blue solid line is the one we use in the main text of the paper. We fix $\sigma_* = 161$ \mbox{km/s} in this plot.}
    \label{bounds tau SIE}
\end{figure*}

The optical depth of the SIE model is subject to the choice of the multiple-image cross section. Across this work we fixed it to be determined by the ``cut region'' in \cite{1994A&A...284..285K}. This region determines the area in which at least two images are formed and it typically encloses the ``caustic region'' where four images are formed. 
Nonetheless, as displayed in Figure 2 of \cite{1994A&A...284..285K}, when the lens axis ratio $q_g$ is small, for example $q_g=0.2$, the caustic region sticks out the cut which increases the overall cross section by a bit. 
Given the fact that most lenses have larger axis ratios, we believe this effect is negligible.

In addition, we also explore the impact of integration limits on the optical depth for the case of SIE lens. 
Figure \ref{bounds tau SIE} shows the comparison using different $\sigma_{\rm min}$ and $\sigma_{\rm max}$. 
In the main text, we set $\sigma_{\rm min} = 0$ \mbox{km/s} and $\sigma_{\rm max} = \infty$ (in practice we set it to a large number, $\rm 10^5 km\ s^{-1}$, around which the number density of the galaxies is approximately equals to 0), which is shown in the sky-blue solid line. 
In reality, the upper and lower velocity dispersion might be different \citep{2007ApJ...658..884C, 2013ApJ...764..184M}, which would have an effect on $\tau (z)$. We tried different choices and, as shown in Figure \ref{bounds tau SIE} with dashed lines, the difference is not very significant when changing the upper and the lower bounds.

\section{BBH merger rate history}
\label{app:Rz}

In this appendix we provide further details about our choices for the merger rate history of BBHs. 
We parametrize the merger rate in the detector frame as:
\begin{equation}
 \frac{d\dot{N}(z)}{dz}\equiv\mathcal{R}_{0} e(z) \frac{dV(z)}{dz}\,,
\end{equation}
where $\mathcal{R}(z=0) \equiv \mathcal{R}_0 = {64.9}_{-33.6}^{+75.5} \Gpc^{-3} \yr^{-1}$ is the local merger rate density \citep{2019ApJ...882L..24A}, and $e(z)$ encapsulates all the redshift information \citep{2011ApJ...739...86Z}. 
The definition of $e(z)$ is then: 

\begin{equation}
\begin{aligned}
e(z) = \frac{\mathcal{R}(z)}{\mathcal{R}_0(1+z)}\,,
\label{eq:ez}
\end{aligned}
\end{equation}
which describes the evolution of the BBH merger relative to the local value. The factor $(1+z)$ converts the source-frame merger rate to the observer-frame merger rate. The numerator $\mathcal{R}(z)$ is given by:
\begin{equation}
\begin{aligned}
\mathcal{R}(z) = \int_{\Delta t_{min}}^{t_{H}(z)}{{\dot{\rho}_{*}(z_{f})\Phi(z_f, \xi)}}P(\Delta t)d\Delta t\,,
\label{eq:rhoc}
\end{aligned}
\end{equation}

\noindent where $z_f$ is the redshift at the binary formation. 
In this expression $\dot{\rho}_{*}(z_{f})$ is the star formation rate, $\Phi(z_f, \xi)$ the metallicity cut and $P(\Delta t)$ the delay time distribution. The delay time $\Delta t$ is the look back time between binary formation and final merger:
\begin{equation}
\begin{aligned}
\Delta t = \int_{z}^{z_f} \frac{dz'}{(1+z')H(z')} \,,
\end{aligned}
\end{equation}
where $H(z)$ is the Hubble rate. We take $P(\Delta t) \propto \frac{1}{\Delta t}$ as the probability distribution of $\Delta t$. This distribution is integrated from a minimum delay time $\Delta t_{\rm min}$ to a maximum one which is equal the age of the Universe at a given redshift $t_{H}(z)$.

\begin{figure*}
\centering
    \includegraphics[height=8cm,width=10cm]{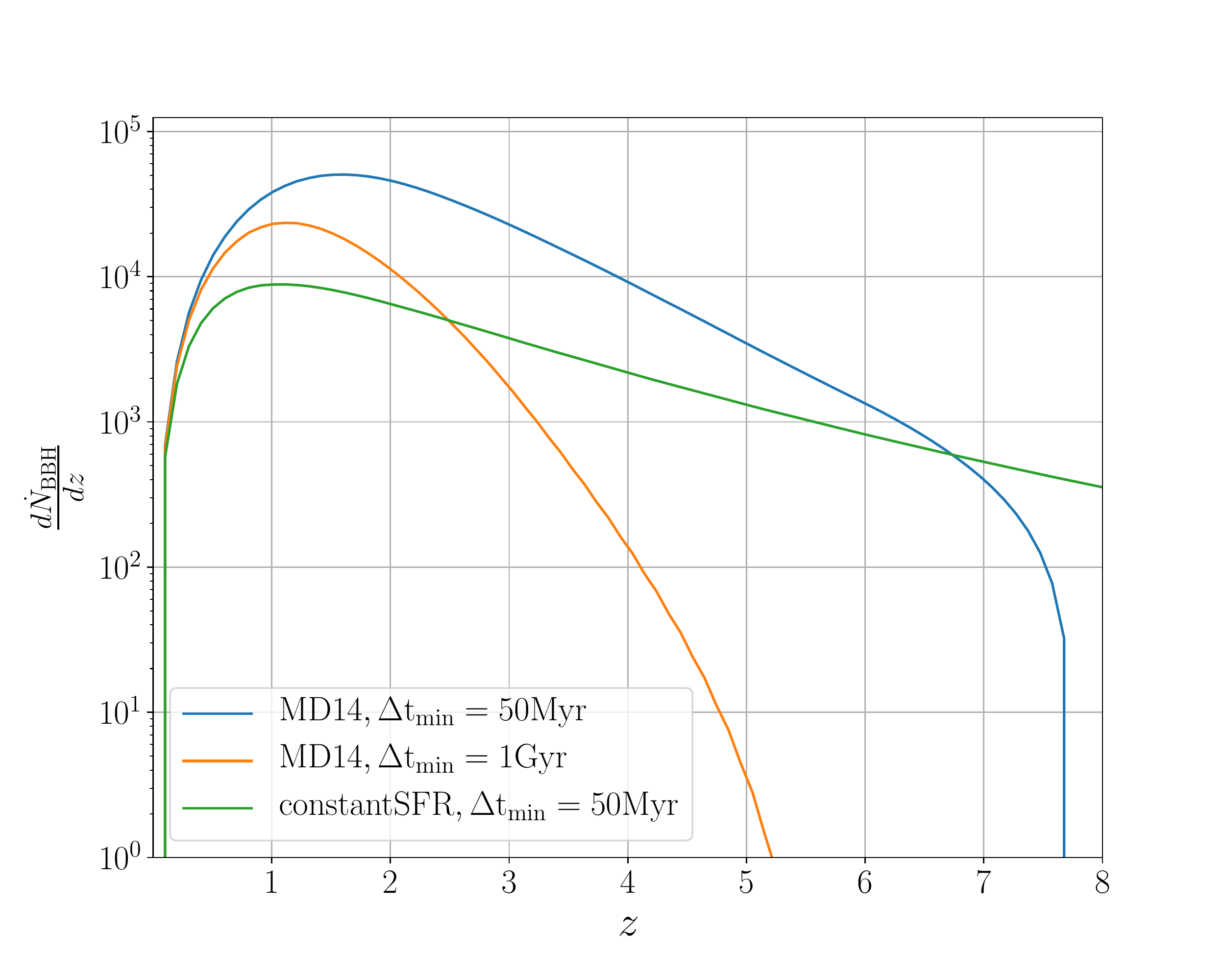}
	\caption{Expected BBH merger rate ($d\dot{N}_{\rm BBH}/dz$) as a function of redshift observed by ET under different assumptions of the star formation rate: SFR model in \citep{2014ARA&A..52..415M} with minimal binary merger delay time $\Delta t_{\rm min} = 50$ Myr (blue line), $\Delta t_{\rm min} = 1$ Gyr (orange line), and constant SFR $\dot{\rho_{*}} = 0.004 M_{\odot}\Mpc^{-3} \rm yr^{-1}$ (green line).}
    \label{fig:ddN/dzdt}
\end{figure*}

The SFR density $\dot{\rho_{*}}$ determines the number of stars that form per unit time and volume. Its unit is $\rm M_{\odot} \Mpc^{-3} \yr^{-1}$. We follow the parametrization of \citet{2014ARA&A..52..415M} (MD14):
\begin{equation}
\begin{aligned}
\dot{\rho_{*}}(z) = 0.015\frac{(1+z)^{2.7}}{1+[(1+z)/2.9]^{5.6}} M_{\odot} \Mpc^{-3} \yr^{-1}\,.
\end{aligned}
\end{equation}
Note that we do not convert $\dot{\rho}_{*}(z_{f})$ from the source frame to the detector frame ${\dot{\rho}_{*}(z_{f})}{/(1+z_f)}$ when integrating Equation \ref{eq:rhoc} \citep{2019ApJ...886L...1V, 2020ApJ...896L..32C}. The binary evolves in its own local frame with its own clock and thus does not have time dilation. However, we do need to convert ${\mathcal{R}(z)}$ from the source frame to the detector frame: $e(z) = \mathcal{R}(z) \times 1/(1+z)$ as in Equation \ref{eq:ez}.  

Since BBH formation favors low metallicity, we include a metallicity dependence factor $\Phi(z_f, \xi)$ which is the fraction of the star formation rate density with metallicity less than $\xi$, where we set $\xi = Z/Z_{\odot} = 0.3$ \citep{2006ApJ...638L..63L}. The final SFR density is thus $\Phi(z_f, \xi = 0.3) \dot{\rho_{*}}(z)$. 

We assume three kinds of SFR models and test how these SFR parameters affect the lensing event rate: (1) MD14 SFR with minimal delay time $\Delta t_{\rm min} = 50$ Myr (the one used in the main text); (2) MD14 SFR with $\Delta t_{\rm min} = 1$ Gyr; (3) a constant SFR model $\dot{\rho_{*}} = 0.004 M_{\odot}\Mpc^{-3} \rm yr^{-1}$ with $\Delta t_{\rm min} = 50$ Myr. 
The $\Delta t_{\rm min}$ is set based on the result in population synthesis studies \citep{2002ApJ...572..407B, 2006LRR.....9....6P, 2013ApJ...779...72D}, and the maximal delay time to $t_{H}(z)$ introduced by the finite age of the Universe. 

The comparison of the observed BBH merger rate $d\dot{N}_{\rm BBH}/dz$ assuming the above 3 SFR scenarios is shown in Figure \ref{fig:ddN/dzdt}, where we use ET as an example for the sensitivity. We can see that when $\Delta t_{\rm min}$ is small or the star formation is more uniform in the cosmic time, the BBH merger events are more extended to higher redshift.

Next we use these different prescriptions for the BBH merger rate to compute the expected rate of lensing. As in the main text we calculate the rate of detecting one strongly lensed image $\dot{N}_{\rm lensing, 1st}$ and two multiply-lensed images $\dot{N}_{\rm lensing, 2nd}$, comparing it with the overall (unlensed) BBH rate $\dot{N}_{\rm BBH}$. 
We summarize our lensing event rate results for the other two SFR scenarios in Table \ref{tab:Rlensing_summary2}.

\begin{table*}
 	\centering
 	\caption{Lensing event rate ($\dot{N}_\text{lensing}$, $\rm yr^{-1}$) assuming $\sigma_* = 161$ \mbox{km/s} for different SFR scenarios. We assume an SIE lens model. The first table assumes constant SFR $\dot{\rho_{*}} = 0.004 M_{\odot}\Mpc^{-3} \yr^{-1}$ and $\Delta t_{\rm min} = 50$ Myr; the second table assumes MD14 \citep{2014ARA&A..52..415M} and $\Delta t_{\rm min} = 1$ Gyr. Similarly, the three columns correspond to $\dot{N}_\text{lensing, 1st}$ and $\dot{N}_\text{lensing, 2nd}$ derived from using $P(\mu)_{\rm 1st}$, $P(\mu)_{\rm 2nd}$ in Equation \ref{eq:Rlensing_mag}, and the expected observed BBH merger event per year $\dot{N}_{\rm BBH}$.}

	\begin{tabular}{lccr} 
		\hline
		  constant SFR, $\Delta t_{\rm min}$ = 50 Myr & $\dot{N}_{\rm lensing, 1st}$ & $\dot{N}_{\rm lensing, 2nd}$ & $\dot{N}_{\rm BBH}$\\
 		\hline
 		aLIGO & 0.09& 0.02& $3.4\times 10^2$ \\ 
 		A+ & 0.58& 0.13 & $1.4 \times 10^3$    \\ 
 		ET &32 &14 & $2.6\times 10^4$   \\ 
 		CE &42 & 31& $3.5\times 10^4$  \\ 
		\hline
	\end{tabular}
	
	\begin{tabular}{lccr} 
		\hline
		  MD 14 SFR, $\Delta t_{\rm min}$ = 1 Gyr & $\dot{N}_{\rm lensing, 1st}$ & $\dot{N}_{\rm lensing, 2nd}$ & $\dot{N}_{\rm BBH}$\\
 		\hline
 		aLIGO & 0.19 & 0.04 & $5.4\times 10^2$ \\ 
 		A+  & 1.02 & 0.3 & $2.7 \times 10^3$  \\ 
 		ET & 12& 8& $3.7\times 10^4$  \\ 
 		CE & 14 & 13 & $4.1\times 10^4$  \\ 
		\hline
	\end{tabular}

\label{tab:Rlensing_summary2}
\end{table*}



\section{Strong lensing event rate from simulations}
\label{app:Rate lensing}

\begin{table*}[b]
 	\centering
 	\caption{We present the lensing event rate of the primary image ($\dot{N}_{\rm lensing, 1st}$), and the lensing event rate with at least 2 images detected($\dot{N}_{\rm lensing, 2nd}$) derived from taking the average and the standard deviation of 100 1-year mock observation samples. The first column is number of the events whose primary images are detected. The second and the third column show the number of the lensing events with at least 2 images detected but with different criterion. The second column compare one random number with the $P(w)$ for both images while the third column compare one random number for each image. These results are consistent with the analytical calculation presented in Table \ref{tab:Rlensing_summary1}.}
	
	\begin{tabular}{lccccr} 
		\hline
		 &  Primary image ($\dot{N}_{\rm lensing, 1st}$) & $\dot{N}_{\rm lensing, 2nd}$ single random number & $\dot{N}_{\rm lensing, 2nd}$ 2 random numbers\\
 		\hline
 		aLIGO & 0.36$\pm$ 0.61 & 0.04$\pm$ 0.19& 0.03$\pm$ 0.17\\ 
 		A+ & 2.87$\pm$ 1.62 & 0.67$\pm$ 0.79  &0.22$\pm$ 0.46 \\ 
 		ET & 94$\pm$ 10 &  50$\pm$ 6 & 45$\pm$ 7\\ 
 		CE & 111$\pm$ 11 & 91$\pm$10.0& 89$\pm$ 10\\ 
		\hline
	\end{tabular}
	\label{tab:Rlensing_summary3}
\end{table*}

%

We present an alternative approach for calculating $\dot{N}_{\rm lensing, 1st}$ and $\dot{N}_{\rm lensing, 2nd}$. 
Instead of solving the integral (Equation \ref{eq:Rlensing_mag}), we now obtain the lensing rate directly from the lensing simulations. Basically we generate 100 mock observation samples as described in Section \ref{subsec: sim} and present the average lensing event rate and the standard deviation of these 100 samples in Table \ref{tab:Rlensing_summary3}. To compute $\dot{N}_{\rm lensing, 1st}$, we draw 1 random number and compare it with $P(w)$ of the primary image. If the random number is smaller than $P(w)$, we consider the image as detected. We use 2 ways to compute $\dot{N}_{\rm lensing, 2nd}$: (1) We generate only 1 random number for each source, and compare it with the $P(w)$ of 
both images. If the random number is smaller than both $P(w)$, we consider the lensing pair is detected as shown in the second column of Table \ref{tab:Rlensing_summary3}; (2) {We generate 1 random number for each image and do the comparison with their own $P(w)$ as shown in the third column of Table \ref{tab:Rlensing_summary3}. Only if both random numbers are smaller than their own $P(w)$, we consider the lensing pair is detected.}

{By comparing the result with Table \ref{tab:Rlensing_summary1} in Section \ref{sec: Rlensing}, we can see that lensing event rate derived from the integration in Equation \ref{eq:Rlensing_mag} can actually give a very good estimation. The estimation of $\dot{N}_{\rm lensing, 1st}$ from the integration in Table \ref{tab:Rlensing_summary1} is very close to the average value from the simulation as shown in the first column of Table \ref{tab:Rlensing_summary3}. While the estimation of $\dot{N}_{\rm lensing, 2nd}$ from the integration is a bit higher than the $\dot{N}_{\rm lensing, 2nd}$ in the third column of Table \ref{tab:Rlensing_summary3} but is consistent with the second column. This is because when doing the integration, we are using the magnification distribution of the secondary image $P(\mu)_{\rm 2nd}$ and thus we only take into account the $P(w)$ of the secondary image. The integral essentially gives the number of the lensing events whose secondary images are detected. Yet in some cases, due to the change in the orientation angle, we might miss the primary image but only detect the secondary one which is less loud. Therefore we think the actual lensing event rate might be lower than the prediction from Equation \ref{eq:Rlensing_mag}. The second method, however, has taken into account these scenarios, and therefore we believe the third column is more realistic. Nevertheless, the second column still gives a reasonable prediction, is less computationally expensive, and can show the dependence of lensing event rate on $\sigma_*$. }


\section{Magnification ratio distribution}
\label{app:mag ratio}
\begin{figure*}
\centering
\includegraphics[height=6cm,width=18cm]{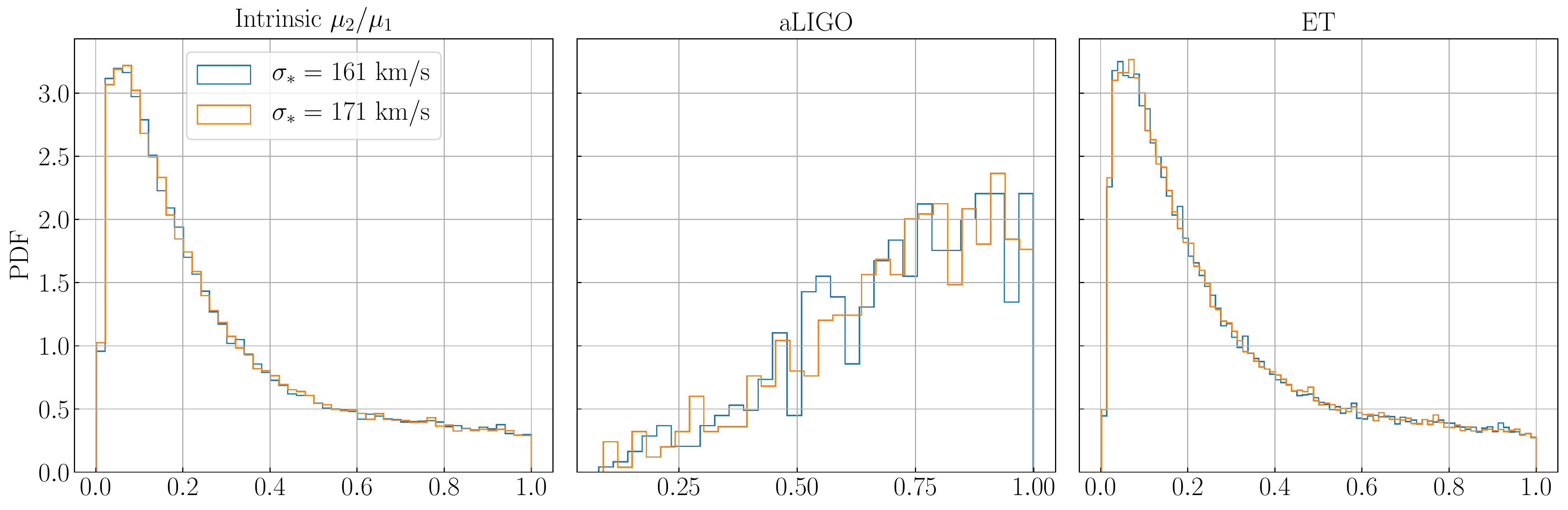}
\includegraphics[height=6cm,width=18cm]{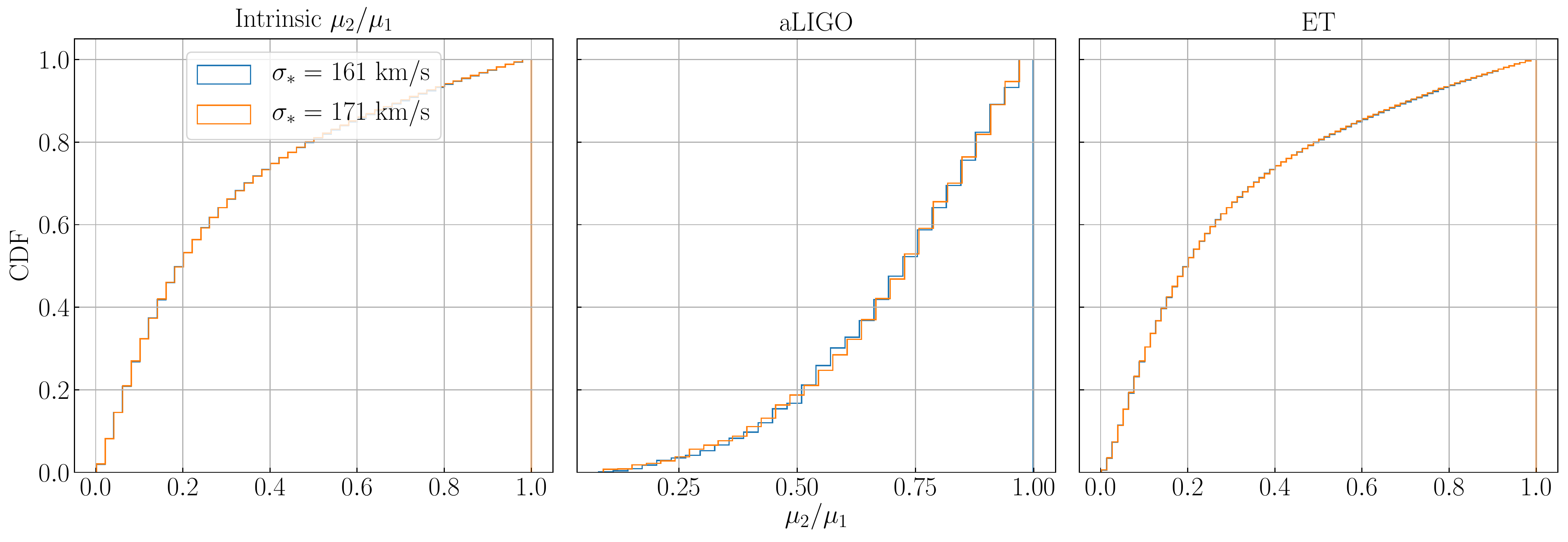}
\caption{Intrinsic and the observed $\mu_2/\mu_1$ ratio. The left, middle, and right panel presents the intrinsic $\mu_2/\mu_1$, $\mu_2/\mu_1$ observed by aLIGO, and $\mu_2/\mu_1$ observed by ET. The upper panels are PDFs and lower panels are CDFs. We demonstrate 2 cases, $\sigma_* = 185 \rm \mbox{km/s}$ and $\sigma_* = 200 \rm \mbox{km/s}$. The magnification ratio is not sensitive to the change in $\sigma_*$.}
\label{fig:mu ratio}
\end{figure*}

In addition to the time delay distribution studied in the main text, the magnification ratio of the secondary over the primary image $\mu_2/\mu_1$ distribution is another potential observable property for GW lensing. According to Equation \ref{eq:mu}, the magnification could potentially probe the axis ratio of the lens galaxies. We show the intrinsic and the observed $\mu_2/\mu_1$ ratio for aLIGO and ET as a demonstration in Figure \ref{fig:mu ratio}. Due to the sensitivity, aLIGO may miss some of the secondary image with small $\mu_2$ and therefore the observed $\mu_2/\mu_1$ distribution peaks near 1. Yet for higher sensitivity like ET, we are able to observe the full distribution. A combined analysis of both time delay distribution and magnification distribution may allow us to simultaneously constrain $\sigma_*$ and $q_g$. 

\begin{figure*}
\centering
\includegraphics[width=\textwidth]{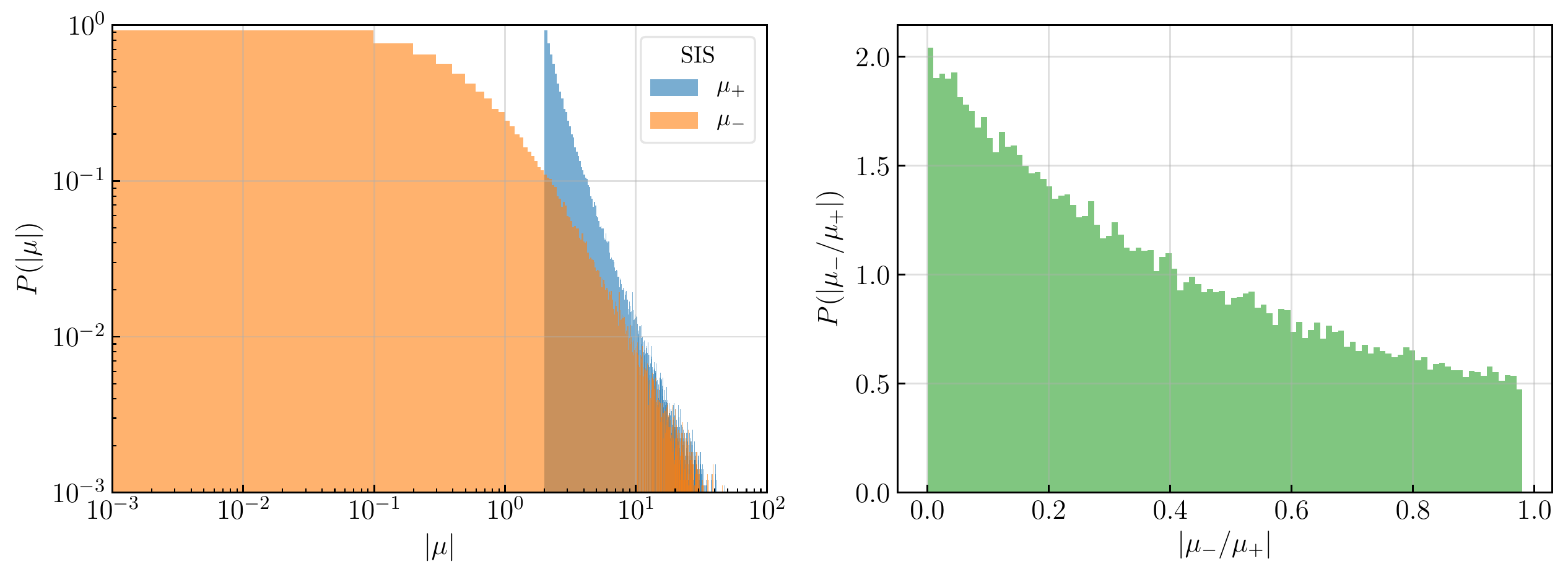}
\caption{{Magnification distribution for the SIS model. Left panel: we present the $P(|\mu|)$ of the individual images. Blue histogram represents the absolute magnification of the primary images, orange represents that of the secondary images. Right panel: we present the magnification ratio.}}
\label{fig:mu_dist_sis}
\end{figure*}

{We can compare these results with the SIS model in which case the magnifications of each of the two images are known analytically: $|\mu_{\pm}|=1\pm1/\beta$, where $\beta$ indicates the angular position of the source \citep{Schneider:1992}. On the left panel of Figure \ref{fig:mu_dist_sis} we present the magnification distribution of each of the images. This plot can be compared to the results in the main text for the SIE model in Figure \ref{fig:Pmu}. Two main differences are noticeable. In the SIS model, the brightest image always has a magnification larger than 2 for angular positions within the Einstein radius. Secondly, the magnification distribution of the parity-odd image, which is always less bright, does not have any peak as in the SIE model. This is because in the SIE model the second brightest image behaves differently when there are 4 images. 
With this information in hand we plot the relative magnification distribution in the right panel of Figure \ref{fig:mu_dist_sis}. 
This can be compared to the upper left panel of Figure \ref{fig:mu ratio}. }

\clearpage
\bibliography{gw_lensing_refs}
\bibliographystyle{aasjournal}

\end{document}